\documentclass[11pt,onecolumn]{IEEEtran}

\usepackage{graphicx}
\usepackage{bm}
\usepackage{amsmath}
\usepackage{amsfonts}
\usepackage{amssymb}

\def \x{\mathbf{x}}

\def \r{\mathbf{r}}

\def \w{\mathbf{w}}

\def \R{\mathbf{R}}
\def \D{\mathbf{D}}
\def \H{\mathbf{H}}

\def \U{\mathbf{U}}

\def \W{\mathbf{W}}

\def \I{\mathbf{I}}

\def \G{\mathbf{G}}

\def \X{\mathbf{X}}

\def \0{\mathbf{0}}

\def \xt {\widetilde{x}}
\def \yt {\widetilde{y}}
\def \tt {\widetilde{t}}

\def \Xt {\widetilde{X}}
\def \Yt {\widetilde{Y}}
\def \xtv {\underline{\widetilde{\x}}}
\def \xv {\underline{\x}}
\def \wv {\underline{\w}}
\def \rv {\underline{\r}}
\def \Xtm {\widetilde{\mathbf{X}}}

\begin{document}

\title{Doppler Resilient Waveforms with Perfect Autocorrelation}

\author{Ali Pezeshki,$^{\star}$~\IEEEmembership{Member,~IEEE,} A. Robert Calderbank,~\IEEEmembership{Fellow,~IEEE,}\\
William Moran,~\IEEEmembership{Member,~IEEE,} and Stephen D.
Howard
\thanks{This work was supported in part by the DARPA
Waveforms for Active Sensing Program under contracts
FA8750-05-2-0285 and N00173-06-1-G006.}
\thanks{A. Pezeshki (corresponding author) and A. R. Calderbank are with The Program in
Applied and Computational Mathematics, Princeton University,
Princeton, NJ 08544, USA (emails: pezeshki@math.princeton.edu;
calderbk@math.princeton.edu).}
\thanks{W. Moran is with the Department of Electrical Engineering
and Computer Science, University of Melbourne, Melbourne 03010,
Australia (e-mail: b.moran@ee.unimelb.edu.au).}
\thanks{S. D.
Howard is with the Defence Science and Technology Organisation,
Edinburgh 05010, Australia.}} \markboth{Pezeshki
\MakeLowercase{\textit{et al.}}: Doppler Resilient Waveforms with
Perfect Autocorrelation}{Pezeshki \MakeLowercase{\textit{et al.}}:
Doppler Resilient Waveforms with Perfect Autocorrelation}

\maketitle

\begin{abstract}

We describe a method of constructing a sequence of phase coded
waveforms with perfect autocorrelation in the presence of Doppler
shift. The constituent waveforms are Golay complementary pairs
which have perfect autocorrelation at zero Doppler but are
sensitive to nonzero Doppler shifts. We extend this construction
to multiple dimensions, in particular to radar polarimetry, where
the two dimensions are realized by orthogonal polarizations. Here
we determine a sequence of two-by-two Alamouti matrices where the
entries involve Golay pairs and for which the sum of the
matrix-valued ambiguity functions vanish at small Doppler shifts.
The Prouhet-Thue-Morse sequence plays a key role in the
construction of Doppler resilient sequences of Golay pairs.

\end{abstract}


\section{Introduction}

The value of perfect autocorrelation sequences in radar imaging is
that their impulse-like autocorrelation function can enable
enhanced range resolution (e.g. see
\cite{Turyn-IT63}--\nocite{Albanese-AES79,Levanon-AES92,Levanon-AES93,Levanon-book}\cite{Golay-IRE61}).
An important class of perfect autocorrelation sequences are
complementary sequences introduced by Golay \cite{Golay-IRE61}.
Golay complementary sequences have the property that the sum of
their autocorrelation functions vanishes at all (integer) delays
other than zero. This means that the sum of the ambiguity
functions (composite ambiguity function) of Golay complementary
sequences is sidelobe free along the zero-Doppler axis, making
them ideal for range imaging.

The concept of complementary sequences was generalized to
multiphase (or polyphase) sequences by Heimiller
\cite{Heimiller-IT61}, Frank \textit{et al.}
\cite{Heimiller-IT62}--\nocite{Frank-IT63}\cite{Frank-IT80}, and
Sivaswami \cite{Sivaswami-IT78}, and to multiple complementary
codes by Tseng and Liu \cite{Tseng-IT72}. Over the past five
decades, the use of complementary and polyphase sequences (and
related codes) have been widely explored for radar imaging, e.g.
see
\cite{Turyn-IT63}--\nocite{Albanese-AES79,Levanon-AES92,Levanon-AES93,Levanon-book,Golay-IRE61,Heimiller-IT61,Heimiller-IT62,Frank-IRT63,Frank-IT80,Sivaswami-IT78,Tseng-IT72,Welti-IT60,Taki-IT69,Kretschmer-AES92,Lewis-AES81,Lewis-AES82}\cite{Lewis-AES83}.
Recently, Deng \cite{Deng-SP04} and Khan \textit{et al.}
\cite{Khan-SPL06} extended the use of polyphase sequences to
orthogonal netted radar (a special case of MIMO radar), and Howard
\textit{et al.} \cite{Howard-Hawaii06} and Calderbank \textit{et
al.} \cite{Calderbank-Asil06} combined Golay complementary
sequences with Alamouti signal processing to enable pulse
compression for multi-channel and fully polarimetric radar
systems. Golay complementary sequences have also been advocated
for the next generation guided radar (GUIDAR) systems
\cite{Harman-AESMag05}.

Despite the attention they received from the radar engineering
community, complementary and polyphase sequences were somewhat
ignored by communication engineers for many years, although their
autocorrelation functions have as low sidelobes as the popular
pseudo noise (PN) sequences. In fact, up until 1990, there were
only a few articles on the use of complementary and polyphase
sequences in communications, among which are the early work by
Reed and Zetterberg \cite{Reed-ComTech64} and the introduction of
orthogonal complementary codes for synchronous spread spectrum
multiuser communications by Suehiro and Hatori
\cite{Suehiro-IT88}. In 1990's, some researchers including
Wilkinson and Jones \cite{Wilkinson-VTC95}, van Nee
\cite{vanNee-Globecom96}, and Ochiai and Imai
\cite{Ochiai-IEICE97} explored the use of Golay complementary
sequences as codewords for OFDM, due to their small peak-to-mean
envelope power ratio (PMEPR). However, the major advances in this
context are due to Davis and Jedwab \cite{Jedwab-IT99} and
Paterson \cite{Paterson-IT00}, who derived tight bounds for the
PMEPR of Golay complementary sequences and related codes from
cosets of the generalized first-order Reed-Muller code.
Construction of low PMEPR codes from cosets of the generalized
first-order Reed-Muller code has also been considered by Schmidt
\cite{Schmidt-IT06} and Schmidt and Finger \cite{Schmidt-WCC05}.
Complementary codes have also been employed as pilot signals for
channel estimation in OFDM systems \cite{Ku-WiComm06}.

Recently, complementary and polyphase codes (in particular
orthogonal complementary codes) have been advocated by Chen
\textit{et al.} \cite{Chen-CommMag01},\cite{Chen-WiCommMag06} and
Tseng and Bell \cite{Bell-Comm00} for enabling interference-free
(both multipath and multi-access) multicarrier CDMA. Other work in
this context include the extension of complementary codes using
the Zadoff-Chu sequence by Lu and Dubey \cite{Dubey-CommLett04}
and cyclic shifted orthogonal complementary codes by Park and Jim
\cite{Park-CommLett06}. In \cite{Li-Comm04}, orthogonal
complementary codes have been used in the design of access-request
packets for contention resolution in random-access wireless
networks.

Despite their many intriguing properties and recent theoretical
advances, in practice a major barrier exists in adoption of
complementary sequences for radar and communications; the perfect
auto-correlation property of these sequences is extremely
sensitive to Doppler shift. Although the shape of the composite
ambiguity function of complementary sequences is ideal along the
zero-Doppler axis, off the zero-Doppler axis it has large
sidelobes in delay, which prevent unambiguous range imaging in
radar or reliable detection in communications. Most
generalizations of complementary sequences, including multiple
complementary sequences and polyphase sequences suffer from the
same problem to some degree. Examples of polyphase sequences that
exhibit some tolerance to Doppler are Frank sequences
\cite{Frank-IT63}, $P1$, $P2$, $P3$, and $P4$ sequences
\cite{Lewis-AES83}, $PX$ sequences \cite{Rapajik-CommLett98}, and
$P(n,k)$ sequences \cite{Felhauer-ELett92},\cite{Felhauer-AES94}.
Sivaswami \cite{Sivaswami-AES82} has also proposed a class of near
complementary codes, called subcomplementary codes, which exhibit
some tolerance to Doppler shift. Subcomplementary codes consist of
a set of $N$ length-$N$ sequences that are phase-modulated by a
binary Hadamard matrix. The necessary and sufficient conditions
for a set of phase-modulated sequences to be subcomplementary have
been derived by Guey and Bell in \cite{Bell-IT98}. The design of
Doppler tolerant polyphase sequences has also been considered for
MIMO radar. In \cite{Khan-SPL06}, Khan \textit{et al.} have used a
harmonic phase structural constraint along with a numerical
optimization method to design a set of polyphase sequences with
resilience to Doppler shifts for orthogonal netted radar.

In this paper, we present a novel and systematic way of designing
a Doppler resilient sequence of Golay complementary waveforms for
radar, for which the composite ambiguity function maintains ideal
shape at small Doppler shifts. The idea is to determine a sequence
of Golay pairs that annihilates the low-order terms of the Taylor
expansion (around zero Doppler) of the composite ambiguity
function. It turns out that the Prouhet-Thue-Morse sequence
\cite{Prouhet-1851}-\nocite{Thue-1906,Thue-19,Morse-1921,
Allouche-book}\cite{Allouche-SETA98} plays a key role in
determining the sequence of Doppler resilient Golay pairs. We then
extend our analysis to the design of a Doppler resilient sequence
of Alamouti waveform matrices of Golay pairs, for which the sum of
the matrix-valued ambiguity functions vanishes at small Doppler
shifts. Alamouti matrices of Golay waveforms have recently been
shown \cite{Howard-Hawaii06},\cite{Calderbank-Asil06} to be useful
for instantaneous radar polarimetry, which has the potential to
significantly increase the performance of fully polarimetric radar
systems, without increasing the receiver signal processing
complexity beyond that of single channel matched filtering. Again,
the Prouhet-Thue-Morse sequence plays a key role in determining
the Doppler resilient sequence of Golay pairs. Finally, numerical
examples are presented to demonstrate the perfect autocorrelation
properties of Doppler resilient Golay pairs at small Doppler
shifts.

\section{Golay Complementary Sequences}

\textbf{Definition 1:} Two length $L$ unimodular sequences of
complex numbers $x[l]$ and $y[l]$ are Golay complementary if the
sum of their autocorrelation functions satisfies
\begin{equation}
\mathrm{corr}_k(x[l])+\mathrm{corr}_k(y[l])=2L\delta_{k,0}, \ \
\text{for} \ k=-(L-1),\cdots,(L-1),
\end{equation}
where $\mathrm{corr}_k(x[l])$ is the autocorrelation of $x[l]$ at
lag $k$ and $\delta_{k,0}$ is the Kronecker delta function.

Let $X(z)=\mathcal{Z}\{x[l]\}$ and $Y(z)=\mathcal{Z}\{y[l]\}$ be
the $z$-transforms of $x[l]$ and $y[l]$ so that
\begin{equation}
\begin{array}{ll}
X(z)&\hspace{-0.2cm}=x[0]+x[1] z^{-1}+\ldots+x[L-1] z^{-(L-1)}\\
Y(z)&\hspace{-0.2cm}=y[0]+y[1] z^{-1}+\ldots+y[L-1] z^{-(L-1)}.
\end{array}
\end{equation}
Then, $x[l]$ and $y[l]$ (or alternatively $X(z)$ and $Y(z)$) are
Golay complementary if $X(z)$ and $Y(z)$ satisfy
\begin{equation}\label{eq:Golayxy1}
X(z)\widetilde{X}(z)+Y(z)\widetilde{Y}(z)=2L
\end{equation}
or equivalently
\begin{equation}\label{eq:Golayxy2}
\|X(z)\|^2+\|Y(z)\|^2=2L,
\end{equation}
where $\widetilde{X}(z)=X^{\ast}(1/z^\ast)$ and
$\widetilde{Y}(z)=Y^{\ast}(1/z^\ast)$ are the $z$-transforms of
$\widetilde{x}[l]=x^\ast[-l]$ and $\widetilde{y}[l]=y^\ast[-l]$,
the time reversed complex conjugates of $x[l]$ and $y[l]$.

Henceforth we drop the discrete time index $l$ from $x[l]$ and
$y[l]$ and simply use $x$ and $y$. We use the notation $(x,y)$
whenever $x$ and $y$ are Golay complementary and call $(x,y)$ a
Golay pair. From \eqref{eq:Golayxy1} it follows that if $(x,y)$ is
a Golay pair then $(\pm x,\pm \widetilde{y})$, $(\pm
\widetilde{x},\pm y)$, and $(\pm \widetilde{x},\pm \widetilde{y})$
are also Golay pairs.

\subsection{Golay Pairs for Radar Detection}\label{sc:GPRD}

Consider a single transmitter/single receiver radar system.
Suppose Golay pairs $(x_0, x_1), (x_2,x_3), \ldots,$
$(x_{N-2},x_{N-1})$ are transmitted over $N$ pulse repetition
intervals (PRIs) to interrogate a radar scene containing a
stationary (relative to the transmitter and receiver) point
target. Let $R_n(z)=\mathcal{Z}\{r_n[k]\}$ denote the
$z$-transform of the radar return associated with the $n$th PRI.
Then, the radar measurement equation can be written (in
$z$-domain) as
\begin{equation}
\begin{array}{cccc}
\underbrace{[R_0(z),\ldots,R_{N-1}(z)]} & =  h
z^{-d_0}\underbrace{[X_{0}(z),\ldots,X_{N-1}(z)]} & \hspace{-.25cm}+ \underbrace{[W_0(z),\ldots,W_{N-1}(z)]}\\
\underline{\r}^T(z) & \ \ \ \ \ \ \ \ \ \underline{\x}^T(z) & \
\underline{\w}^T(z)
\end{array}
\end{equation}
where the delay $d_0$ in $z^{-d_0}$ depends on the target range
$r_0$ and is given by $d_0=[2r_0/(c t_0)]$, where $t_0$ is the
``chip'' interval (time interval between two consecutive values in
$x[l]$ or $y[l]$), $c$ is the speed of light, and $[a]$ denotes
the integer part of $a$. Without loss of generality, from hereon
we assume that $d_0=0$, centering the delay axis at the target
location. The scalar $h$ is the target scattering coefficient,
which we assume to be proper complex normal with zero mean and
variance $2\sigma_h^2$, but fixed over the $N$ RPIs. Elements of
$\underline{\w}^T(z)$ are $z$-transforms of iid samples of proper
complex white Gaussian noises with variance $2\sigma_w^2$.

If we process the radar return vector $\underline{\r}^{T}(z)$ by a
receiver vector of the form
\begin{equation}
\underline{\widetilde{\x}}(z)=[\widetilde{X}_0(z),\ldots,\widetilde{X}_{N-1}(z)]^T
\end{equation}
then the receiver output will be
\begin{equation}\label{eq:MFout1}
\begin{array}{ll}
U(z)=\r^T(z)\xtv(z)&\hspace{-.2cm}=h
\xv^T(z)\xtv(z)+\wv^T(z)\xtv(z)\vspace{0.2cm}\\
&\hspace{-0.2cm}= NL h + \wv^T(z)\xtv(z),
\end{array}
\end{equation}
where the second equality follows by replacing $\xv^T(z)\xtv(z)$
with
\begin{equation}\label{eq:sumamb}
\xv^T(z)\xtv(z)=(\|X_0(z)\|^2+\|X_1(z)\|^2)+\ldots+(\|X_{N-2}(z)\|^2+\|X_{N-1}(z)\|^2)=NL.
\end{equation}

The term $\xv^T(z)\xtv(z)$ is the $z$-transform of the composite
ambiguity function of Golay pairs $(x_0, x_1),\ldots,$
$(x_{N-2},x_{N-1})$ along the zero-Doppler axis. We notice that
$\xv^T(z)\xtv(z)$ is a constant, which means that the composite
ambiguity function of $(x_0, x_1), \ldots, (x_{N-2},x_{N-1})$
vanishes at all (integer) delays along the zero-Doppler axis.

Transforming \eqref{eq:MFout1} back to the time domain, we have
\begin{equation}
u[k]=\mathcal{Z}^{-1}\{U(z)\}=NL h \delta_{k,0}+n[k],
\end{equation}
where $n[k]$ is a proper complex white Gaussian noise with
variance $2\sigma_n^2=(NL)2\sigma_w^2$. This shows that detecting
a stationary point in range amounts to the following Gaussian
hypothesis test
\begin{equation}\label{eq:HP2by2}
u[k]=\left\{\begin{array}{*{20}l}
n \sim CN[0,2 \sigma_n^2 ] & \hspace{.3cm}: \mathrm{H_0}\vspace{.2cm}\\
NLh+n \sim CN[0,(2 N^2L^2\sigma_h^2+ 2 \sigma_n^2)] &
\hspace{.3cm}: \mathrm{H_1}
\end{array}\right.
\end{equation}
where $CN[0,2 \sigma_n^2 ]$ denotes the proper complex normal
distribution with mean zero and variance $2\sigma_n^2$.

\textit{Remark 1:} In the above analysis, the radar return
associated with each PRI is processed separately at the receiver,
that is each radar return is correlated with its corresponding
waveform and then all the correlator outputs are added together.
Hence the receiver output (in time domain) is
\begin{equation}\label{eq:Remark1}
\begin{array}{ll}
u[k]&\hspace{-.2cm}=\sum\limits_{n=0}^{N-1}\mathrm{xcorr}_k(r_n[k'],x_n[k'])\vspace{.2cm}\\
&\hspace{-.2cm}=h\sum\limits_{n=0}^{N-1}\mathrm{corr}_k(x_n[k'])+n[k]\vspace{.2cm}\\
&\hspace{-.2cm}=NL h \delta_{k,0}+n[k],
\end{array}
\end{equation}
where $\mathrm{xcorr}_k(r_n[k'],x_n[k'])$ is the cross-correlation
between $r_n[k']$ and $x_n[k']$ at lag $k$. If we want to process
all the PRIs together then we must correlate the augmented radar
return $r_a[k]$,
\begin{equation}
r_a[k]=r_0[k]+r_1[k-D]+\ldots+r_{N-1}[k-(N-1)D]
\end{equation}
with the augmented waveform $x_a[k]$,
\begin{equation}
x_a[k]=x_0[k]+x_1[k-D]+\ldots+x_{N-1}[k-(N-1)D],
\end{equation}
where $D$ is the delay associated with a PRI. The receiver output
in this case is
\begin{equation}
\begin{array}{ll}
u_a[k]&\hspace{-.2cm}=\mathrm{xcorr}_k(r_a[k'],x_a[k'])\vspace{.4cm}\\
&\hspace{-.2cm}=h\sum\limits_{n=0}^{N-1}\mathrm{corr}_k(x_n[k'])
+h\mathop{\sum\limits_{n'=0}^{N-1}\sum\limits_{n''=0}^{N-1}}\limits_{n'\neq
n''}\mathrm{xcorr}_{k}(x_{n'}[k'-n'D],x_{n''}[k'-n''D])+n_a[k]\vspace{.2cm}\\
&\hspace{-.2cm}=NL h \delta_{k,0}
+h\mathop{\sum\limits_{n'=0}^{N-1}\sum\limits_{n''=0}^{N-1}}\limits_{n'\neq
n''}\mathrm{xcorr}_{k}(x_{n'}[k'-n'D],x_{n''}[k'-n''D])+n_a[k],
\end{array}
\end{equation}
where $n_a[k]$ is a noise term. The cross terms
$\mathrm{xcorr}_{k}(x_{n'}[k'-n'D],x_{n''}[k'-n''D])$ result in
range sidelobes whose peaks are offset by integer multiples of $D$
from the origin $k=0$. Thus, by processing each radar return
separately as in \eqref{eq:Remark1} we can avoid range sidelobes
caused by cross-correlations between different waveforms. However,
the Doppler resolution will be limited by the time duration of a
single waveform, whereas in the case where all the returns are
processed together the Doppler resolution is enhanced due to
having a longer transmit pulse.

\subsection{Effect of Doppler on Golay Pairs}

We now consider the case where the target moves at a constant
speed, causing a Doppler shift of $\theta$ [rad] between two
consecutive PRIs. We assume that the radar PRI is short enough
that during the $N$ PRIs where the Golay pairs are transmitted the
target range remains approximately the same. Then the composite
radar measurement is given by
\begin{equation}\label{eq:RetDopp}
\rv^T(z,\theta)=  h \xv^T(z) \D(\theta) + \wv^T(z),
\end{equation}
where $\D(\theta)$ is the following diagonal Doppler modulation
matrix:
\begin{equation}\label{eq:Dmat}
\D(\theta)=\mathrm{diag}(1,e^{j\theta},\ldots,e^{j(N-1)\theta}).
\end{equation}

If we now process the radar measurement vector $\rv^T(z,\theta)$
using the receiver vector $\xtv(z)$ the receiver output will be
\begin{equation}
U(z,\theta)=\rv^T(z,\theta)\xtv(z)=h G(z,\theta)+\wv^T(z)\xtv(z),
\end{equation}
where $G(z,\theta)$ is the $z$-transform of the composite
ambiguity function of $(x_0, x_1),\ldots,(x_{N-2},x_{N-1})$, and
is given by
\begin{equation}\label{eq:Gz}
G(z,\theta)=\xv^T(z)\D(\theta)\xtv(z)=\|X_{0}(z)\|^2+e^{j\theta}\|X_{1}(z)\|^2+\ldots+e^{j(N-1)\theta}\|X_{N-1}(z)\|^2.
\end{equation}
We notice that off the zero-Doppler axis ($\theta\neq 0$) the
composite ambiguity function $G(z,\theta)$ is not sidelobe-free at
integer delays. In fact, even small Doppler shifts can result in
large sidelobes at integer delays.

One way to solve this problem is to use a bank of Doppler filters
to estimate the unknown Doppler shift $\theta$ and then compensate
for the Doppler effect by post-multiplying \eqref{eq:RetDopp} by
$\D^H(\theta)$ (where $H$ denotes Hermitian transpose) prior to
applying $\xtv(z)$. However, since even a slight mismatch in
Doppler can result in large sidelobes, we have to cover the
possible Doppler range at a fine resolution, which requires the
use of many Doppler filters. This motivates the question of
whether it is possible to design \textit{Doppler resilient} Golay
pairs $(x_0, x_1), \ldots, (x_{N-2},x_{N-1})$ so that
$G(z,\theta)=\sum_{n=0}^{N-1}e^{j n\theta}\|X_n(z)\|^2\approx
\alpha z^0$, where $\alpha$ is constant, for a reasonable range of
Doppler shifts $\theta$. We are looking to construct the Golay
pairs $(x_0, x_1), \ldots, (x_{N-2},x_{N-1})$ so that
$G(z,\theta)$ (which is a two-sided polynomial of degree $L-1$ in
$z^{-1}$) vanishes at every delay but zero.

\section{Doppler Resilient Golay Pairs}\label{sc:1by1}

In this section we consider the design of Doppler resilient
sequences of Golay pairs. More precisely, we describe how to
select Golay pairs $(x_0, x_1), \ldots, (x_{N-2},x_{N-1})$ so that
in the Taylor expansion of $G(z,\theta)$ around $\theta=0$ the
coefficients of all terms up to a certain order, say $M$, vanish
at all nonzero delays.

Consider the Taylor expansion of $G(z,\theta)$ around $\theta=0$,
i.e.,
\begin{equation}\label{eq:Taylor}
G(z, \theta)=\sum\limits_{m=0}^{\infty}C_m(z)\theta^m,
\end{equation}
where
\begin{equation}
C_m(z)=\sum_{n=0}^{N-1} n^m \|X_n(z)\|^2, \ \ \ \ \ \text{for} \ \
m=0,1,2,3,\ldots
\end{equation}
In general, the coefficients $C_m(z)$, $m=1,2,3,\ldots$ are
two-sided polynomials in $z^{-1}$ of the form
\begin{equation}\label{eq:Cmpoly}
C_m(z)=\sum_{l={-(L-1)}}^{L-1}c_{m,l} z^{-l}, \ \ \ \ \ \ \
m=1,2,3,\ldots
\end{equation}
For instance, the first coefficient $C_1(z)$ is
\begin{equation}
C_1(z)=0 \|X_0(z)\|^2+1 \|X_1(z)\|^2+2 \|X_2(z)\|^2+\ldots+(N-1)
\|X_{N-1}(z)\|^2.
\end{equation}
Noting that $(x_0,x_1), \ldots, (x_{N-2},x_{N-1})$ are Golay pairs
we can simplify $C_1(z)$ as
\begin{equation}
C_1(z)=N(N-2) L/2+ \|X_1(z)\|^2+ \|X_3(z)\|^2+\ldots+
\|X_{N-1}(z)\|^2.
\end{equation}
Each term of the form
$\|X_{2k+1}(z)\|^2=X_{2k+1}(z)X_{2k+1}^\ast(1/z^\ast)$ is a
two-sided polynomial of degree $L-1$ in the delay operator
$z^{-1}$, which can not be matched with any of the other terms, as
we have already taken into account all the Golay pairs.
Consequently, $C_1(z)$ is a two-sided polynomial in $z^{-1}$ of
the form $C_1(z)=\sum_{l={-(L-1)}}^{L-1}c_{1,l} z^{-l}$.

We wish to design the Golay pairs $(x_0, x_1), \ldots,
(x_{N-2},x_{N-1})$ so that $c_{1,l}$ vanish for all nonzero $l$.
More generally, we wish to design $(x_0, x_1), \ldots,
(x_{N-2},x_{N-1})$ so that in the Taylor expansion in
\eqref{eq:Taylor} the coefficients of all the terms up to a given
order $M$ vanish at all nonzero delays, i.e. $c_{m,l}=0$, for all
$m$ ($1\le m \le M$) and for all nonzero $l$. Although not
necessary, we continue to carry the term $0^m \|X_0(z)\|^2$ in
writing $C_m(z)$ for reasons that will become clear. We note that
there is no need to consider the zero-order term, as $C_0(z)=NL$.

\subsection{The Requirement that $C_1(z)$ Vanish at All Nonzero Delays}

To provide intuition, we first consider the case $N=2^2=4$, where
Golay pairs $(x_0,x_1)$ and $(x_2,x_3)$ are transmitted over four
PRIs. Then, as the following calculation shows, $C_1(z)$ will
vanish at all nonzero delays if the Golay pairs $(x_0,x_1)$ and
$(x_2,x_3)$ are selected such that $(x_1,x_3)$ is also a Golay
pair:
\begin{equation}\label{eq:C1N4}
\begin{array}{ll}
C_1(z)=& \hspace{-0.2cm}\underbrace{0\|X_0(z)\|^2+\|X_1(z)\|^2}+\underbrace{2\|X_2(z)\|^2+3\|X_3(z)\|^2}\vspace{0.05cm}\\
& \hspace{-0.2cm}\underbrace{\ \ \ \ \ \ \ 1\|X_1(z)\|^2 \ \ \ \ \ \ \ \ \ \ \ \ \ \ \ 2\times 2L+1\|X_3(z)\|^2}\vspace{0.05cm}\\
& \hspace{-0.2cm} \ \ \ \ \ \ \ \ \ \ \ \ \ \ \ \ \ \ \ \ \ \ \ \
\  3\times 2L
\end{array}
\end{equation}
The trick is to break $3$ into $2+1$, and then pair the extra
$\|X_3(z)\|^2$  with $\|X_1(z)\|^2$. Note that it is easy to
choose the pairs $(x_0,x_1)$ and $(x_2,x_3)$ such that $(x_1,x_3)$
is also a Golay pair. For example, let $(x,y)$ be an arbitrary
Golay pair, then $(x_0=x,x_1=y)$, $(x_2=-\yt,x_3=\xt)$, and
$(x_1=y,x_3=\xt)$ are all Golay pairs. Other combinations of $\pm
x$, $\pm \xt$, $\pm y$, and $\pm \yt$ are also possible. The
calculation in \eqref{eq:C1N4} shows that it is possible to make
$C_1(z)$ ($M=1$) vanish at all nonzero delays with $N=2^{1+1}$
Golay sequences $x_0,\ldots,x_3$.

\subsection{The Requirement that $C_1(z)$ and $C_2(z)$ Vanish at All Nonzero Delays}

It is easy to see that when $N=4$ it is not possible to force
$C_2(z)$ ($M=2$) to zero at all nonzero delays. However, this is
possible when $N=2^{2+1}=8$. As the following calculations show,
we can make both $C_1(z)$ and $C_2(z)$ vanish at all nonzero $l$
if we select the Golay pairs $(x_0,x_1), \ldots, (x_6,x_7)$ such
that $(x_1,x_3)$, $(x_5,x_7)$, and $(x_3,x_7)$ are also Golay
pairs.\footnote{In writing \eqref{eq:C11} and \eqref{eq:C21} we
have dropped the argument $z$ for simplicity.}

\textit{Making $C_1(z)$ vanish:}
\begin{equation}\label{eq:C11}
\begin{array}{l}
C_1(z)=\\
\underbrace{0\|X_0\|^2+1\|X_1\|^2}\ + \ \underbrace{2\|X_2\|^2+3\|X_3\|^2} \ + \ \underbrace{4\|X_4\|^2+5\|X_5\|^2} \ + \ \underbrace{6\|X_6\|^2+7\|X_7\|^2}\\
 \ \ \ \ \ \ \ \ \ \ \ \ \ \ \ \ \ \ \ \ \ \ \ \ \
\ \ \ \ \ \ \ 2\times 2L+ \ \ \ \ \ \ \ \ \ \ \ \ \ \ \ \ 4\times
2L+ \
\ \ \ \ \ \ \ \ \ \ \ \ \ \ 6\times 2L+\\
\underbrace{[(1-0)=1]\|X_1\|^2 \ \ \ [(3-2)=1]\|X_3\|^2} \ \ \
\underbrace{\ [(5-4)=1]\|X_5\|^2  \ \ \ [(7-6)=1]\|X_7\|^2}\\ \ \
\ \ \ \ \ \ \ \ \ \ \ \ \ \ \ \ \ \ \ 3\times 2L \ \ \ \ \ \ \ \ \
\ \ \ \ \ \ \ \ \ \ \ \ \ \ \ \ \ \ \ \ \ \  \ \ \ \ \ \ \ \ \ \ \
\ 11\times 2L
\end{array}
\end{equation}

\textit{Making $C_2(z)$ vanish:}
\begin{equation}\label{eq:C21}
\begin{array}{l}
C_2(z)=\\
\underbrace{0^2\|X_0\|^2+1^2\|X_1\|^2} \ + \ \underbrace{2^2\|X_2\|^2+3^2\|X_3\|^2} \ + \ \underbrace{4^2\|X_4\|^2+5^2\|X_5\|^2} \ + \ \underbrace{6^2\|X_6\|^2+7^2\|X_7\|^2}\\
\ \ \ \ \ \ \ \ \ \ \ \ \ \ \ \ \ \ \ \ \ \ \ \ \ \ \ \ \ \ \ \ \
\ \ \ 4\times 2L+ \ \ \ \ \ \ \  \ \ \ \ \ \ \ \ \ 16\times 2L+ \
\ \ \ \ \ \ \ \ \ \ \ \ \ \ \ \ 36\times 2L+\\
\underbrace{[(1^2-0^2)=1]\|X_1\|^2 \ \ [(3^2-2^2)=5]\|X_3\|^2} \ \
\ \underbrace{ [(5^2-4^2)=9]\|X_5\|^2  \ \ \
[(7^2-6^2)=13]\|X_7\|^2}\\
 \ \ \ \ \ \ \ \ \ \ \ \ \ \ \ \ \ \ \ \ \ \ 5\times 2L+ \ \ \ \
\ \ \ \ \ \ \ \ \ \ \ \ \ \ \ \ \ \ \ \ \ \ \ \ \ \ \  \ \ \ \ \ \ \ \ \ \ \ \ \ \ 61 \times 2L+\\
 \underbrace{\ \ \ \ \ \ \ [(3^2-2^2-1^2+0^2)=4]\|X_3\|^2
\ \ \ \ \ \ \ \ \ \ \ \ \ \ \ \ \ \ \ \ \ [(7^2-6^2-5^2+4^2)=4]\|X_7\|^2 \ \ \ \ \ \ \ \ }\\
\ \ \ \ \ \ \ \ \ \ \ \ \ \ \ \ \ \ \ \ \ \ \ \ \ \ \ \ \ \ \ \
[(0^2+1^2+2^2+\ldots+7^2)=70]\times 2L
\end{array}
\end{equation}

Note that it is easy to select the Golay pairs $(x_0,x_1), \ldots,
(x_6,x_7)$ such that $(x_1,x_3)$, $(x_5,x_7)$, and $(x_3,x_7)$ are
also Golay pairs. For example, $(x_0=x,x_1=y)$,
$(x_2=-\yt,x_3=\xt)$, $(x_4=-\yt,x_5=\xt)$, and $(x_6=x,x_7=y)$,
where $(x,y)$ is an arbitrary Golay pair, satisfy all the extra
Golay pair conditions.

We notice that what allows us to make both $C_1(z)$ and $C_2(z)$
vanish at all nonzero $l$ is the identity
\begin{equation}
3^m-2^m-1^m+0^m=7^m-6^m-5^m+4^m, \ \ \ \text{for} \ \ m=1,2
\end{equation}
or alternatively
\begin{equation}\label{eq:part1}
(0^m+3^m+5^m+6^m)-(1^m+2^m+4^m+7^m)=0, \ \ \ \text{for} \ \ m=1,2,
\end{equation}
where $m=1$ and $m=2$ correspond to the calculations for $C_1(z)$
and $C_2(z)$, respectively. In other words, the reason $C_1(z)$
and $C_2(z)$ can be forced to zero at all nonzero delays is that
the set $\mathbb{S}=\{0,1,\ldots,7\}$ can be partitioned into two
disjoint subsets $\mathbb{S}_0=\{0,3,5,6\}$ and
$\mathbb{S}_1=\{1,2,4,7\}$ whose elements satisfy
\eqref{eq:part1}. This is a special case of the Prouhet (or
Prouhet-Tarry-Escott) problem
\cite{Allouche-SETA98},\cite{Lahmer-ScriptaMath47} which we will
discuss in more detail later in this section. But for now we just
note that $\mathbb{S}_{0}$ is the set of all numbers in
$\mathbb{S}$ that correspond to the zeros in the length-$8$
Prouhet-Thue-Morse sequence (PTM)
\cite{Prouhet-1851}-\nocite{Thue-1906,Thue-19,Morse-1921,
Allouche-book}\cite{Allouche-SETA98}
\begin{equation}
(s_k)_{k=0}^7 \ = \ 0 \ 1 \ 1 \ 0 \ 1 \ 0 \ 0 \ 1,
\end{equation}
and $\mathbb{S}_{1}$ is the set of all numbers in $\mathbb{S}$
that correspond to the ones in $(s_k)_{k=0}^7$.

A key observation here is that the extra Golay pair conditions we
had to introduce to make $C_1(z)$ and $C_2(z)$ vanish at all
nonzero $l$ are all associated with pairs of the form $(x_p,x_q)$
where $p$ and $q$ are odd, and $p\in \mathbb{S}_0$ and $q\in
\mathbb{S}_1$. This suggests a close connection between the
Prouhet-Thue-Morse sequence and the way Golay sequences $x_0, x_1,
\ldots, x_{N-1}$ must be paired.

\subsection{The Requirement that $C_1(z)$ Through $C_M(z)$ Vanish}

We now address the general problem of selecting the Golay pairs
$(x_0,x_1), \ldots, (x_{N-2},x_{N-1})$ to make $C_m(z)$,
$m=1,2,\ldots, M$ vanish at all nonzero delays. We begin with some
definitions and results related to the Prouhet-Thue-Morse
sequence.

\textbf{Definition
2.}\cite{Prouhet-1851}-\nocite{Thue-1906,Thue-19,Morse-1921,
Allouche-book}\cite{Allouche-SETA98} The Prouhet-Thue-Morse (PTM)
sequence $\mathcal{S}=(s_k)_{k\ge 0}$ over $\{0,1\}$ is defined by
the following recursions:
\begin{enumerate}
\item $s_0=0$ \item $s_{2k}=s_{k}$ \item
$s_{2k+1}=\overline{s}_{k}=1-s_{k}$
\end{enumerate}
for all $k>0$, where $\overline{s}=1-s$ denotes the binary
complement of $s\in\{0,1\}$.

For example, the PTM sequence of length 32 is
\begin{equation}
\mathcal{S}=(s_k)_{k=0}^{31} \ = \ 0 \ 1 \ 1 \ 0 \ 1 \ 0 \ 0 \ 1 \
1 \ 0 \ 0 \
 1 \ 0 \ 1 \ 1 \ 0 \ 1 \ 0 \ 0 \ 1 \ 0 \ 1 \ 1 \ 0 \ 0 \ 1 \ 1 \ 0 \ 1 \ 0 \ 0 \
 1.
\end{equation}

\textbf{Prouhet's
problem.}\cite{Allouche-SETA98},\cite{Lahmer-ScriptaMath47} Let
$\mathbb{S}=\{0,1,\ldots, N-1\}$ be the set of all integers
between $0$ and $N-1$. The Prouhet's problem (or
Prouhet-Tarry-Escott problem) is the following. Given $M$, is it
possible to partition $\mathbb{S}$ into two disjoint subsets
$\mathbb{S}_0$ and $\mathbb{S}_1$ such that $\sum\limits_{p\in
\mathbb{S}_{0}} p^{m} = \sum\limits_{q\in \mathbb{S}_{1}} q^{m}$
for all $0\le m\le M$? Prouhet's proved that this is possible when
$N=2^{M+1}$ and that the partitions are identified by the PTM
sequence.

\textbf{Theorem 1
(Prouhet).}\cite{Allouche-SETA98},\cite{Lahmer-ScriptaMath47} Let
$\mathcal{S}=(s_k)_{k\ge 0}$ be the PTM sequence. Define
\begin{equation}
\begin{array}{ll}
\mathbb{S}_0 & =\{p\in \mathbb{S}=\{0,1,2,\ldots, 2^{M+1}-1\}| \ s_p=0\}\vspace{0.2cm}\\
\mathbb{S}_1 & =\{q\in \mathbb{S}=\{0,1,2,\ldots, 2^{M+1}-1\}| \
s_q=1\}
\end{array}
\end{equation}
Then, for any $m$ with $0\le m \le M$ we have
\begin{equation}
\sum\limits_{p\in \mathbb{S}_{0}} p^{m} = \sum\limits_{q\in
\mathbb{S}_{1}} q^{m}.
\end{equation}

\textbf{Lemma 1.} Let $(x_0,x_1), \ldots, (x_{N-2},x_{N-1})$,
$N=2^{M+1}$ be Golay pairs. Let $\mathbb{X}_{0}=\{x_p | p\in
\mathbb{S}_{0}\}$ and $\mathbb{X}_{1}=\{x_q | q\in
\mathbb{S}_{q}\}$. Then, neither $\mathbb{X}_0$ nor $\mathbb{X}_1$
contains any of the Golay pairs $(x_0,x_1), \ldots,
(x_{N-2},x_{N-1})$.

\textit{Proof:} The Golay pairs $(x_0,x_1), \ldots,
(x_{N-2},x_{N-1})$ are of the form $(x_{2k},x_{2k+1})$, where
$k=0,1,\ldots,$ $N/2-1$. From the definition of the PTM sequence
we have $s_{2k+1}=\overline{s}_{k}=\overline{s}_{2k}$. Therefore,
$x_{2k}$ and $x_{2k+1}$ cannot be in the same set.$\square$

\textbf{Lemma 2.} Assume that the Golay pairs $(x_0,x_1)$,
$\ldots$, $(x_{N-2},x_{N-1})$, $N=2^{M+1}$ are such that all pairs
of the form $(x_{2k'+1}\in \mathbb{X}_{0},x_{2k''+1}\in
\mathbb{X}_{1})$, i.e., all pairs of the form
$(x_{2k'+1},x_{2k''+1})$ with $2k'+1\in\mathbb{S}_0$ and
$2k''+1\in\mathbb{S}_{1}$, are also Golay complementary. Then,
\begin{equation}
\|X_p(z)\|^2=\|X_{p'}(z)\|^2 \ \ \ \text{and} \ \ \
\|X_q(z)\|^2=\|X_{q'}(z)\|^2
\end{equation}
for all $p,p'\in \mathbb{S}_{0}$ (i.e. for all $x_p,x_{p'}\in
\mathbb{X}_0$) and for all $q,q'\in \mathbb{S}_{1}$ (i.e. for all
$x_q,x_{q'}\in \mathbb{X}_1$), and all pairs of the form $(x_p\in
\mathbb{X}_0,x_q\in \mathbb{X}_1)$, i.e. all pairs $(x_p,x_q)$
with $p\in \mathbb{S}_0$ and $q\in \mathbb{S}_{1}$, are Golay
complementary.

\textit{Proof:} Assume $p=2k$ is even and $p\in \mathbb{S}_{0}$.
Then $q=2k+1$ is odd and $q\in \mathbb{S}_1$. We know that the
pair $(x_{p=2k}\in\mathbb{X}_{0},x_{q=2k+1}\in\mathbb{X}_1)$ is
Golay complementary, as all the original Golay pairs
$(x_0,x_1),\ldots,(x_{N-2},x_{N-1})$ are of the form
$(x_{2k},x_{2k+1})$, hence
\begin{equation}\label{eq:pair1}
\|X_p(z)\|^2+\|X_q(z)\|^2=2L.
\end{equation}
Let $p'\in \mathbb{S}_0$ and assume $p'$ is odd. Then, since
$q=2k+1\in \mathbb{S}_1$ and all pairs of the form $(x_{2k'+1}\in
\mathbb{X}_{0},x_{2k''+1}\in \mathbb{X}_{1})$ are Golay
complementary (from our assumption), we have
\begin{equation}\label{eq:pair2}
\|X_{p'}(z)\|^2+\|X_q(z)\|^2=2L.
\end{equation}
Subtracting \eqref{eq:pair2} from \eqref{eq:pair1} gives
\begin{equation}\label{eq:corrpp}
\|X_p(z)\|^2=\|X_{p'}(z)\|^2.
\end{equation}
Since \eqref{eq:corrpp} is true for any even $p\in \mathbb{S}_0$
and any odd $p'\in \mathbb{S}_0$ it must be true for any $p,p'\in
\mathbb{S}_{0}$, or equivalently any $x_p,x_{p'}\in \mathbb{X}_0$.
Similarly, we can prove that $\|X_q(z)\|^2=\|X_{q'}(z)\|^2$ for
all $x_q,x_{q'}\in \mathbb{X}_{1}$. Since at least one element
from $\mathbb{X}_0$ forms a pair with one element in
$\mathbb{X}_1$ (e.g. $(x_0,x_1)$) then all pairs of the form
$(x_p\in \mathbb{X}_0,x_q\in \mathbb{X}_1)$ must be Golay
complementary.$\square$

\textit{Remark 2:} We note that to construct Golay pairs
$(x_0,x_1), \ldots, (x_{N-2},x_{N-1})$, $N=2^{M+1}$ that satisfy
the conditions of Lemma 2 we can consider an arbitrary Golay pair
$(x,y)$ and then arbitrarily choose $x_p\in \mathbb{X}_0$ from the
set $\{x,-x,\xt,-\xt\}$ and $x_q\in \mathbb{X}_1$ from the set
$\{y,-y,\yt,-\yt\}$, for any $p\in \mathbb{S}_0$ and any $q\in
\mathbb{S}_1$.

We now present the main result of this section by stating the
following theorem.

\textbf{Theorem 2.} The coefficients $C_1(z),\ldots,C_M(z)$ in the
Taylor expansion \eqref{eq:Taylor} will vanish at all nonzero
delays if the Golay pairs $(x_0,x_1), \ldots, (x_{N-2},x_{N-1})$,
$N=2^{M+1}$ are selected such that all pairs $(x_p,x_q)$ where $p$
and $q$ are odd and $p\in \mathbb{S}_{0}$ and $q\in
\mathbb{S}_{1}$ are also Golay complementary.

\textit{Proof:} From Lemma 2, we have
$\|X_p(z)\|^2=\|X_{p'}(z)\|^2$ for all $p,p'\in \mathbb{S}_0$ and
$\|X_q(z)\|^2=\|X_{q'}(z)\|^2$ for all $q,q'\in \mathbb{S}_1$.
Therefore, we can write $C_m(z)$ ($1\le m\le M)$ as
\begin{equation}\label{eq:proof1}
C_m(z)=\sum\limits_{n=0}^{N-1}n^{m}
\|X_n(z)\|^2=(\sum\limits_{p\in
\mathbb{S}_{0}}p^m)\|X_0(z)\|^2+(\sum\limits_{q\in
\mathbb{S}_{1}}q^m)\|X_1(z)\|^2.
\end{equation}
From the Prouhet theorem (Theorem 1), we have $\sum\limits_{p\in
\mathbb{S}_{0}}p^m=\sum\limits_{q\in \mathbb{S}_{1}}q^m=\beta$,
where $\beta$ is constant. Therefore, we have
\begin{equation}
C_m(z)=\beta(\|X_0(z)\|^2+\|X_1(z)\|^2)= 2\beta L .
\end{equation}

\section{Doppler Resilient Golay Pairs for Fully Polarimetric Radar
Systems}\label{sc:2by2}

Fully polarimetric radar systems are capable of simultaneously
transmitting and receiving on two orthogonal polarizations. The
use of two orthogonal polarizations increases the degrees of
freedom and can result in significant improvement in detection
performance. Recently, Howard \textit{et al.}
\cite{Howard-Hawaii06} (also see \cite{Calderbank-Asil06})
proposed a novel approach to radar polarimetry that uses
orthogonal polarization modes to provide essentially independent
channels for viewing a target, and achieve diversity gain. Unlike
conventional radar polarimetry, where polarized waveforms are
transmitted sequentially and processed non-coherently, the
approach in \cite{Howard-Hawaii06} allows for
\textit{instantaneous radar polarimetry}, where polarization modes
are combined coherently on a pulse by pulse basis. Instantaneous
radar polarimetry enables detection based on full polarimetric
properties of the target and hence can provide better
discrimination against clutter. When compared to a radar system
with a singly-polarized transmitter and a singly-polarized
receiver the instantaneous radar polarimetry can achieve the same
detection performance (same false alarm and detection
probabilities) with a substantially smaller transmit energy, or
alternatively it can detect at substantially greater ranges for a
given transmit energy \cite{Howard-Hawaii06}.

A key ingredient of the approach in \cite{Howard-Hawaii06} is a
unitary Alamouti matrix of Golay waveforms that has a perfect
\textit{matrix-valued ambiguity function} along the zero-Doppler
axis. The unitary property of the waveform matrix allows for
detection in range based on the full polarimetric properties of
the target, without increasing the receiver signal processing
complexity beyond that of single channel matched filtering. We
show in this section that it is possible to design a sequence of
Alamouti matrices of Golay waveforms, for which the sum of the
matrix-valued ambiguity functions vanishes at all nonzero
(integer) delays for small Doppler shifts.

\begin{figure}[!tp]
\begin{center}
\begin{tabular}{c}
\includegraphics[width=2.5in]{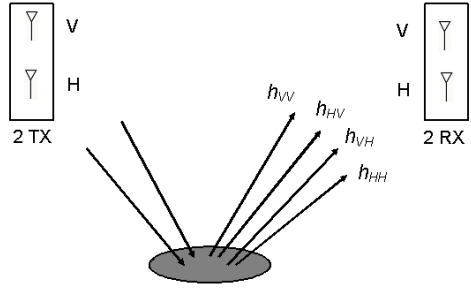}
\end{tabular}
\end{center}
\caption{Scattering model for a fully polarimetric radar system,
with a dually-polarized transmit and a dually-polarized receive
antenna.}\label{f:2by2r}
\end{figure}

Figure \ref{f:2by2r} shows the scattering model of the fully
polarimetric radar system considered in \cite{Howard-Hawaii06},
where $h_{VH}$ denotes the scattering coefficient into the
vertical polarization channel from a horizontally polarized
incident field. Howard \textit{et al.} employ Alamouti signal
processing \cite{Alamouti-JSAC98} to coordinate the transmission
of $(N/2)$ Golay pairs $(x_{0},x_1), \ldots, (x_{N-2},x_{N-1})$
over vertical and horizontal polarizations during $N$ PRIs. The
waveform matrix is of the form
\begin{equation}
\X(z)=\begin{pmatrix}X_{0}(z) & -\Xt_{1}(z) & \ldots & X_{2k}(z) &
-\Xt_{2k+1}(z)
& \ldots & X_{N-2}(z) & -\Xt_{N-1}(z)\\
X_{1}(z) & \Xt_{0}(z) & \ldots & X_{2k+1}(z) & \Xt_{2k}(z) &
\ldots & X_{N-1}(z) & \Xt_{N-2}(z)
\end{pmatrix},
\end{equation}
where different rows in $\X(z)$ correspond to vertical and
horizontal polarizations, and different columns correspond to
different time slots (PRIs).

The radar measurement matrix $\R(z)$ for this transmission scheme
is given by
\begin{equation}
\R(z)=\H\X(z)\D(\theta)+\W(z),
\end{equation}
where $\H$ is the 2 by 2 target scattering matrix, with entries
$h_{VV}$, $h_{VH}$, $h_{HV}$, and $h_{HH}$, $\W(z)$ is a 2 by $N$
noise matrix with entries that are iid proper complex normal with
zero mean and variance $2\sigma_w^2$, and $\D(\theta)$ is the
diagonal Doppler modulation matrix introduced in \eqref{eq:Dmat}.

If we process $\R(z)$ with a receiver matrix $\Xtm(z)$ of the form
\begin{equation}
\Xtm(z)=\begin{pmatrix} \Xt_{0}(z) & -X_1(z) & \ldots &
\Xt_{2k}(z) & -X_{2k+1}(z) & \ldots & \Xt_{N-2}(z) & -X_{N-1}(z)\\
\Xt_{1}(z) & X_0(z) & \ldots & \Xt_{2k+1}(z) & X_{2k}(z) & \ldots
& \Xt_{N-1}(z) & X_{N-2}(z)
\end{pmatrix}^{T}
\end{equation}
then the receiver output will be
\begin{equation}
\U(z,\theta)=\R(z)\Xtm(z)=\H\G(z,\theta)+\W(z)\Xtm(z),
\end{equation}
where the term $\G(z,\theta)=\X(z)\D(\theta)\Xtm(z)$ can be viewed
as the $z$-transform of a \textit{matrix-valued ambiguity
function} for $\X(z)$. Along the zero-Doppler axis, where
$\D(\theta=0)=\I$, due to the interplay between Alamouti signal
processing and the Golay property, the term $\G(z,\theta)$ reduces
to
\begin{equation}\label{eq:Golaymat}
\G(z,0)=\X(z)\Xtm(z)=\begin{pmatrix}
\sum\limits_{n=0}^{N-1}\|X_n(z)\|^2=NL \hfill &
\sum\limits_{k=0}^{N/2-1}
(1-1)X_{2k}(z)\Xt_{2k+1}(z)=0 \vspace{0.3cm}\\
\sum\limits_{k=0}^{N/2-1} (1-1)\Xt_{2k}(z)X_{2k+1}(z)=0 & \hfill
 \sum\limits_{n=0}^{N-1}\|X_n(z)\|^2=NL
\end{pmatrix}.
\end{equation}
This shows that $\X(z)$ has a perfect matrix-valued ambiguity
function along the zero-Doppler axis; that is along the
zero-Doppler axis $G(z,\theta)$ vanishes at all nonzero (integer)
delays, and is unitary at zero-delay. A consequence of
\eqref{eq:Golaymat} is that detecting a point target in range
reduces to a simple Gaussian hypothesis test, for which the
likelihood ratio detector is the same as an energy detector.
However, off the zero-Doppler axis the property in
\eqref{eq:Golaymat} no longer holds, and the elements of the
matrix-valued ambiguity function $G(z,\theta)$ can have large
sidelobes, even at small Doppler shifts.

We consider how the Golay pairs $(x_{0},x_1), \ldots,
(x_{N-2},x_{N-1})$ must be selected so that for small Doppler
shifts we have
\begin{equation}\label{eq:CGz}
\G(z,\theta)=\X(z)\D(\theta)\Xtm(z)=\begin{pmatrix}
G_{1}(z,\theta) & G_2(z,\theta) \\ \widetilde{G}_{2}(z,\theta) &
G_1(z,\theta)
\end{pmatrix}\approx\begin{pmatrix} NL & 0\\
0 & NL
\end{pmatrix},
\end{equation}
where
\begin{equation}
\begin{array}{ll}
G_1(z,\theta)&\hspace{-0.2cm}=\sum\limits_{n=0}^{N-1}e^{j n \theta}\|X_n(z)\|^2\vspace{0.2cm}\\
&\hspace{-0.2cm}=\|X_0(z)\|^2+e^{j\theta}\|X_1(z)\|^2+\ldots+e^{j(N-1)\theta}\|X_{N-1}(z)\|^2
\end{array}
\end{equation}
and
\begin{equation}
\begin{array}{ll}
G_2(z,\theta)&\hspace{-0.2cm}=\sum\limits_{k=0}^{N/2-1} (e^{j
2k\theta}-e^{j(2k+1)\theta})X_{2k}(z)\Xt_{2k+1}(z)\vspace{0.2cm}\\
&\hspace{-0.2cm}=(1-e^{j\theta})X_{0}(z)\Xt_{1}(z)+\ldots+(e^{j(N-2)\theta}-e^{j(N-1)\theta})X_{N-2}(z)\Xt_{N-1}(z).
\end{array}
\end{equation}

The diagonal term of $\G(z,\theta)$, i.e., $G_1(z,\theta)$, is
equal to the single channel composite ambiguity function
$G(z,\theta)$ in \eqref{eq:Gz}. Therefore, we can use Theorem 2 to
design the Golay pairs $(x_{0},x_1), \ldots, (x_{N-2},x_{N-1})$,
$N=2^{M+1}$ such that in the Taylor expansion \eqref{eq:Taylor}
the coefficients $C_m(z)$, $m=1,2,\ldots, M$ vanish at all nonzero
delays. Thus, from now on we only discuss how the off-diagonal
term $G_2(z,\theta)$ can be forced to zero for small Doppler
shifts.

Consider the Taylor expansion of $G_2(z,\theta)$ around
$\theta=0$, i.e.,
\begin{equation}\label{eq:G2Taylor}
G_2(z,\theta)=\sum\limits_{m=0}^{\infty}B_m(z)\theta^m,
\end{equation}
where
\begin{equation}
B_m(z)=\sum\limits_{k=0}^{N/2-1}((2k)^m-(2k+1)^m)
X_{2k}(z)\Xt_{2k+1}(z), \ \ \ \ \text{for} \ \ m=0,1,2,\ldots .
\end{equation}
In general, the coefficients $B_m(z)$, $m=1,2,3,\ldots$ are
two-sided polynomials in $z^{-1}$ of the form
\begin{equation}\label{eq:Bmpoly}
B_m(z)=\sum_{l={-(L-1)}}^{L-1} b_{m,l} z^{-l}, \ \ \ \ \ \ \
m=1,2,3,\ldots
\end{equation}
For instance, the first coefficient $B_1(z)$ is
\begin{equation}\label{eq:B1z}
B_1(z)=(0-1)
X_0(z)\Xt_1(z)+\ldots+((N-2)-(N-1))X_{N-2}(z)\Xt_{N-1}(z).
\end{equation}
Each term of the form $X_{2k}(z)\Xt_{2k+1}(z)$ in \eqref{eq:B1z}
is a two-sided polynomial of degree $L-1$ in $z^{-1}$, and since
in general the terms $X_{2k}(z)\Xt_{2k+1}(z)$ for different values
of $k$ do not cancel each other, $B_1(z)$ is also a two-sided
polynomial of degree $L-1$ in $z^{-1}$.

Suppose that the Golay pairs $(x_{0},x_1), \ldots,
(x_{N-2},x_{N-1})$, $N=2^{M+1}$ satisfy the conditions of
Theorem~2 so that $C_1(z), \ldots, C_M(z)$ vanish at all nonzero
delays. We wish to determine the extra conditions required for
$(x_{0},x_1), \ldots, (x_{N-2},x_{N-1})$ to force $B_1(z), \ldots,
B_M(z)$ to zero at all delays. As we show, again the PTM sequence
is the key to finding the zero-forcing conditions. The zero-order
term $B_0(z)$ is always zero and hence we do not consider it in
our discussion.

\subsection{The Requirement that $B_1(z)$ Vanish}

Again, to gain intuition, we first consider the case $N=2^2=4$.
Then, as the following calculation shows, $B_1(z)$ will vanish if
the Golay pairs $(x_0,x_1)$ and $(x_2,x_3)$ are selected so that
$X_0(z)\Xt_1(z)=-X_2(z)\Xt_3(z)$:
\begin{equation}
\begin{array}{ll}
B_1(z)=&\hspace{-0.2cm}(0-1)X_0(z)\Xt_1(z)+\underbrace{(2-3)X_2(z)\Xt_3(z)}\\
&\hspace{-0.2cm}\underbrace{\ \ \ \ \ \ \ \ \ \ \ \ \ \ \ \ \ \ \
\ \ \ \ \ \
-(2-3)X_0(z)\Xt_1(z)}\\
& \hspace{-0.2cm} \ \ \ \ \ \ \ \ \ \ \ \ \ \ \ \  \ \ \ \ \ \ \ \
0
\end{array}
\end{equation}

In summary, to make $C_1(z)$ vanish at all nonzero delays and to
force $B_1(z)$ to zero at the same time, the Golay pairs
$(x_0,x_1)$ and $(x_2,x_3)$ must be selected such that $(x_1,x_3)$
is also a Golay pair and $X_0(z)\Xt_1(z)=-X_2(z)\Xt_3(z)$. If we
let $(x,y)$ be an arbitrary Golay pair then it is easy to see that
$(x_0=x,x_1=y)$, $(x_2=-\yt,x_3=\xt)$ satisfy these conditions
(other choices are also possible). The Alamouti waveform matrix
$\X(z)$ for this choice of Golay pairs is given by
\begin{equation}
\X(z)=\begin{pmatrix} X_0(z)=X(z) & -\Xt_1(z)=-\Yt(z) &
X_2(z)=-\Yt(z) & -\Xt_3(z)=-X(z)
\\ X_1(z)=Y(z) & \Xt_0(z)=\Xt(z) & X_3(z)=\Xt(z) & \Xt_2(z)=-Y(z) \end{pmatrix}.
\end{equation}

\subsection{The Requirement that $B_1(z)$ and $ B_{2}(z)$ Vanish:}

Let us now consider the case $N=2^3=8$. Then, as the following
calculations show, both $B_1(z)$ and $ B_{2}(z)$ will vanish if we
select $(x_0,x_1),\ldots,(x_6,x_7)$ such that
$X_0(z)\Xt_1(z)=-X_2(z)\Xt_3(z)=-X_4(z)\Xt_5(z)=X_6(z)\Xt_7(z)$.

\textit{Making $B_1(z)$ vanish:}
\begin{equation}\label{eq:B1Van}
\begin{array}{ll}
B_1(z)=\\
\ \ \ \ \underbrace{(0-1)X_0\Xt_1} \ \ \ \ + \ \ \ \ \ \underbrace{(2-3)X_2\Xt_3} \ \ \ \ +  \ \ \ \underbrace{(4-5)X_4\Xt_5} \ \ \ \ + \ \ \ \ \underbrace{(6-7)X_6\Xt_7}\\
\underbrace{[(0-1)=-1]X_0\Xt_1 \ \ [-(2-3)=1]X_0\Xt_1 \ \
[-(4-5)=1]X_0\Xt_1 \ \ [(6-7)=-1]X_0\Xt_1}\\ \ \ \ \ \ \ \ \ \ \ \
\ \ \ \ \ \ \ \ \ \ \ \ \ \ \ \ \ \ \ \ \ \ \ \  \ \ \ \ \ \ \ \ \
\ \  \ \ \ \ \ \  0
\end{array}
\end{equation}

\textit{Making $B_2(z)$ vanish:}
\begin{equation}\label{eq:B2Van}
\begin{array}{ll}
B_2(z)=\\
\ \ \ \ \underbrace{(0^2-1^2)X_0\Xt_1} \ \ \ \ \ + \ \ \ \ \ \ \underbrace{(2^2-3^2)X_2\Xt_3} \ \  +  \ \ \underbrace{(4^2-5^2)X_4\Xt_5} \ \ \ \ \ + \ \ \ \ \underbrace{(6^2-7^2)X_6\Xt_7}\\
\underbrace{[(0^2-1^2)=-1]X_0\Xt_1 \ \  [-(2^2-3^2)=5]X_0\Xt_1} \
\ \underbrace{[-(4^2-5^2)=9]X_0\Xt_1 \ \
[(6^2-7^2)=-13]X_0\Xt_1}\\
\underbrace{\ \ \ \ \ \ \ \ \ \ \ \ [0^2-1^2-2^2+3^2=4]X_0\Xt_1 \
\ \ \ \ \ \ \ \ \ \ \ \ \ \ \ \ \ \ \
\ [-4^2+5^2+6^2-7^2=-4] X_0\Xt_1 \ \ \ \ \ \ \ \ \ } \\
\ \ \ \ \ \ \ \ \ \ \ \ \ \ \ \ \ \ \ \ \ \ \ \ \ \ \ \ \ \ \ \ \
\ \  \ \ \ \ \ \ \ \ \ \ \  \ \ \ \ \ \ \ \ \ \ \  0
\end{array}
\end{equation}
In writing \eqref{eq:B1Van} and \eqref{eq:B2Van} we have dropped
the argument $z$ for simplicity.

In summary, to make $C_1(z)$ and $C_2(z)$ vanish at all nonzero
delays and to force $B_1(z)$ and $B_2(z)$ to zero at the same
time, the Golay pairs $(x_0,x_1), \ldots, (x_6,x_7)$ must satisfy
the conditions of Theorem 2, and the within-pair cross-spectral
densities must satisfy
\begin{equation}\label{eq:crosscond}
X_0(z)\Xt_1(z)=-X_2(z)\Xt_3(z)=-X_4(z)\Xt_5(z)=X_6(z)\Xt_7(z).
\end{equation}
It is easy to see that the Golay pairs in the following waveform
matrix $\X(z)$ satisfy all these conditions:
\begin{equation}\label{eq:Xz2m}
\X(z)=\begin{pmatrix} \X_0(z) & \X_1(z) & \X_1(z) & \X_0(z)
\end{pmatrix},
\end{equation}
where $\X_0(z)$ and $\X_1(z)$ are given by
\begin{equation}\label{eq:X0X1}
\begin{array}{ccc}
\X_0(z)=\begin{pmatrix} X(z) & -\Yt(z) \\ Y(z) &
\Xt(z)\end{pmatrix} & \text{and} & \X_1(z)=\begin{pmatrix} -\Yt(z)
& -X(z) \\ \Xt(z) & -Y(z)\end{pmatrix},
\end{array}
\end{equation}
and $(x,y)$ is an arbitrary Golay pair.

The trick in forcing $B_1(z)$ and $B_2(z)$ to zero is to cleverly
select the signs of the cross-correlation functions
(cross-spectral densities) between the two sequences in every
Golay pair relative to the cross-correlation function
(cross-spectral density) for $x_0$ and $x_1$. If we let $0$ and
$1$ correspond to the positive and negative signs respectively, we
observe that the sequence of signs in \eqref{eq:crosscond}
corresponds to the length-$4$ PTM sequence. In the next section,
we show that the PTM sequence is in fact the right sequence for
specifying the relative signs of the cross-correlation functions
between the Golay sequences in each Golay pair.

\textit{Remark 3:} Representing $\X_0(z)$ and $\X_1(z)$ by $0$ and
$1$ respectively, we notice that the placements of $\X_0(z)$ and
$\X_1(z)$ in $\X(z)$ are also determined by the length-$4$ PTM
sequence.

\subsection{The Requirement that $B_1(z)$ Through $B_M(z)$ Vanish}

We now consider the general case $N=2^{M+1}$ where Golay pairs
$(x_0,x_1)$, $\ldots$, $(x_{N-2},x_{N-1})$ are used to construct a
Doppler resilient waveform matrix $\X(z)$. We have the following
theorem.

\textbf{Theorem 3:} Let $N=2^{M+1}$ and let $(x_0,x_1)$, $\ldots$,
$(x_{N-2},x_{N-1})$ be Golay pairs. Then, for any $m$ between $1$
and $M$, $B_m(z)$ will vanish at all delays if
\begin{equation}
X_{2k}(z)\Xt_{2k+1}(z)=(-1)^{s_k}X_{0}(z)\Xt_{1}(z), \ \ \ \ \
\text{for all $k$}, \ 0\le k \le N/2-1,
\end{equation}
where $s_k$ is the $k$th element in the PTM sequence.

\textit{Proof:} For any $m$ ($1\le m \le M$), $B_m(z)$ may be
written as
\begin{equation}\label{eq:Bmzproof}
\begin{array}{ll}
B_m(z)&\hspace{-0.2cm}=\sum\limits_{k=0}^{N/2-1} ((2k)^m-(2k+1)^m)
X_{2k}(z)\Xt_{2k+1}(z)\vspace{0.2cm}\\
&\hspace{-0.2cm}=\left[\sum\limits_{k=0}^{N/2-1}
(-1)^{s_k}((2k)^m-(2k+1)^m)\right] X_0(z)\Xt_1(z),
\end{array}
\end{equation}
where the second equality in \eqref{eq:Bmzproof} follows by
replacing $X_{2k}(z)\Xt_{2k+1}(z)$ with
$(-1)^{s_k}X_0(z)\Xt_1(z)$. Since in the PTM sequence
$s_k=s_{2k}=\overline{s}_{2k+1}$, we can rewrite
\eqref{eq:Bmzproof} as
\begin{equation}\label{eq:Bmzproof1}
\begin{array}{ll}
B_m(z)&\hspace{-0.2cm}=\left[\sum\limits_{k=0}^{N/2-1}
(-1)^{s_{2k}}(2k)^m-(-1)^{\overline{s}_{2k+1}}(2k+1)^m \right]
X_0(z)\Xt_1(z)\vspace{0.2cm}\\
&\hspace{-0.2cm}=\left[\sum\limits_{k=0}^{N/2-1}
(-1)^{s_{2k}}(2k)^m+(-1)^{s_{2k+1}}(2k+1)^m\right] X_0(z)\Xt_1(z)\vspace{0.2cm}\\
&\hspace{-0.2cm}=\left[\sum\limits_{k=0}^{N-1}
(-1)^{s_{k}}k^m\right]X_0(z)\Xt_1(z).
\end{array}
\end{equation}
However, from the Prouhet theorem (Theorem 1), it is easy to see
that $\sum\limits_{k=0}^{N-1}(-1)^{s_k} k^m=0$, and therefore
$B_m(z)=0$.$\square$

Finally, we note that it is always possible to find Golay pairs
$(x_0,x_1), \ldots, (x_{N-2},x_{N-1})$ that satisfy the conditions
of both Theorem 2 and Theorem 3. Suppose $(x_0,x_1), \ldots,
(x_{N-2},x_{N-1})$ are built from an arbitrary Golay pair $(x,y)$
(as explained in Section \ref{sc:1by1}) to satisfy the conditions
of Theorem 2. Then, we can apply the time reverse operator and
change the sign of the elements within the pairs to satisfy the
conditions of Theorem 3, as the Golay property is invariant to
time reversal and changes in the signs of the Golay sequences
within a pair.

\section{Numerical Examples}

In this section, we present numerical examples to verify the
results of Sections \ref{sc:1by1} and \ref{sc:2by2} and compare
our Doppler resilient design to a conventional scheme, where the
same Golay pair is repeated.

\subsection{Single Channel Radar System}

We first consider the case of a single channel radar system. In
this case, the composite ambiguity function $G(z,\theta)$ is given
by \eqref{eq:Gz} and has a Taylor expansion of the form
\eqref{eq:Taylor}. Following Theorem 2, we coordinate the
transmission of eight Golay pairs $(x_0,x_1), \ldots,
(x_{14},x_{15})$ over $N=16$ PRIs to make the Taylor expansion
coefficients $C_1(z), \ldots, C_3(z)$ ($M=3$) vanish at all
nonzero delays. Starting from a Golay pair $(x,y)$, it is easy to
verify that the eight Golay pairs in the following waveform vector
$\xv^T(z)$ satisfy the conditions of Theorem 2:
\begin{equation}\label{eq:X1vec}
\xv^T(z)=\begin{pmatrix} \xv_0^T(z) & \xv_1^T(z) & \xv_1^T(z) &
\xv_0^T(z) & \xv_1^T(z) & \xv_0^T(z) & \xv_0^T(z) & \xv_1^T(z)
\end{pmatrix},
\end{equation}
where $\xv_0^T(z)=[X(z) \ \ \ Y(z)]$ and $\xv_1^T(z)=[-\Yt(z) \ \
\ \Xt(z)]$.

\textit{Remark 4:} Representing $\xv_0^T(z)$ and $\xv_1^T(z)$ by
$0$ and $1$ respectively, we notice that the placements of
$\xv_0^T(z)$ and $\xv_1^T(z)$ in $\xv^T(z)$ are determined by the
length-$8$ PTM sequence.

We compare the Doppler resilient transmission scheme in
\eqref{eq:X1vec} with a conventional transmission scheme, where
the same Golay pair $(x_0=x,x_1=y)$ is transmitted during all
PRIs, resulting in a waveform vector $\xv_c^T(z)$ of the form
\begin{equation}\label{eq:X0vec}
\xv_c^T(z)=\begin{pmatrix} \xv_0^T(z) & \xv_0^T(z) & \xv_0^T(z) &
\xv_0^T(z) & \xv_0^T(z) & \xv_0^T(z) & \xv_0^T(z) & \xv_0^T(z)
\end{pmatrix},
\end{equation}
with the composite ambiguity function
\begin{equation}
\begin{array}{ll}
G_c(z,\theta)&\hspace{-.2cm}=\xv_c^T(z)\D(\theta)\xtv_c(z)\vspace{.2cm}\\
&\hspace{-.2cm}=\|X_0(z)\|^2+e^{j\theta}\|X_1(z)\|^2+\ldots+e^{j(N-2)\theta}\|X_0(z)\|^2+e^{j(N-1)\theta}\|X_1(z)\|^2.
\end{array}
\end{equation}

The pair $(x,y)$ used in constructing $\xv^T(z)$ and $\xv_c^T(z)$
can be any Golay pair. Here, we choose $(x,y)$ to be the following
length-8 ($L=8$) Golay pair:
\begin{equation}\label{eq:basexy}
\begin{array}{llll}
x[l]&\hspace{-.2cm}= \{1,1,-1,1,1,1,1,-1\} & \Longleftrightarrow & X(z)=1+z^{-1}-z^{-2}+z^{-3}+z^{-4}+z^{-5}+z^{-6}-z^{-7} \vspace{0.2cm}\\
y[l]&\hspace{-.2cm}= \{-1,-1,1,-1,1,1,1,-1\} & \Longleftrightarrow
& Y(z)=-1-z^{-1}+z^{-2}-z^{-3}+z^{-4}+z^{-5}+z^{-6}-z^{-7}
\end{array}
\end{equation}

\begin{figure}[!htp]
\begin{center}
\begin{tabular}{ccc}
\includegraphics[height=1.3in]{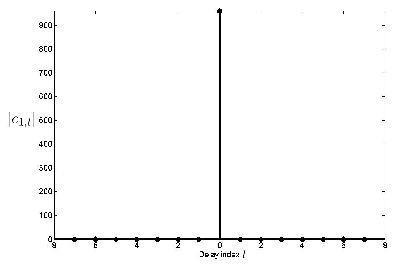}&
\includegraphics[height=1.3in]{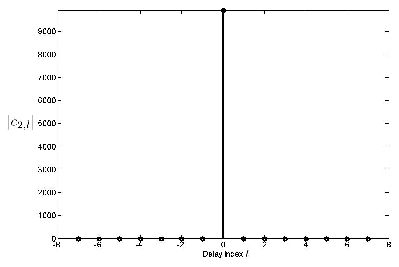}&
\includegraphics[height=1.3in]{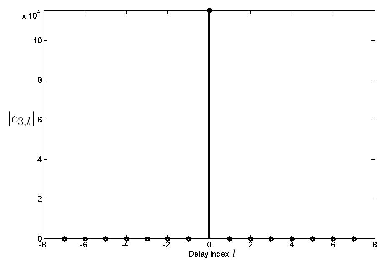}\\
(a) $|c_{1,l}|$ & (b) $|c_{2,l}|$ & (c) $|c_{3,l}|$
\end{tabular}
\end{center}
\caption{The plots of the magnitudes of the coefficients $c_{m,l}$
of two-sided polynomials $C_m(z)=\sum\limits_{l=-(L-1)}^{L-1}
c_{m,l}z^{-l}$, $m=1,2,3$ versus delay index $l$: (a) $|c_{1,l}|$,
(b) $|c_{2,l}|$, and (c) $|c_{3,l}|$.}\label{f:diag}
\end{figure}

Referring to the Taylor expansion of $G(z,\theta)$ in
\eqref{eq:Taylor}, the coefficients $C_1(z)$, $C_2(z)$, and
$C_3(z)$ are each two-sided polynomials of degree $L-1=7$ in
$z^{-1}$ of the form \eqref{eq:Cmpoly}. Figures
\ref{f:diag}(a)-(c) show the plots of the magnitudes of the
coefficients $c_{m,l}$, $m=1,2,3$ of these polynomials versus
delay index $l$. The plots show that $C_1(z)$, $C_2(z)$, and
$C_3(z)$ indeed vanish at all nonzero delays.

\begin{figure}[!htp]
\begin{center}
\begin{tabular}{cc}
\includegraphics[height=2in]{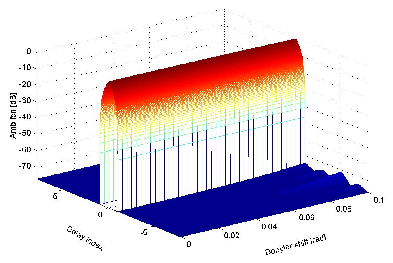}&
\includegraphics[height=2in]{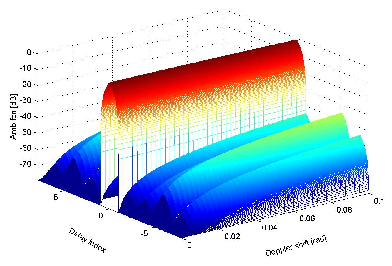}\\
(a) $G(z,\theta)$ & (b) $G_c(z,\theta)$
\end{tabular}
\end{center}
\caption{(a) The plot of the composite ambiguity function
$G(z,\theta)$ (corresponding to the Doppler resilient transmission
scheme) versus delay index $l$ and Doppler shift $\theta$, (b) the
plot of the composite ambiguity function $G_c(z,\theta)$
(corresponding to the conventional transmission scheme) versus
delay index $l$ and Doppler shift $\theta$.}\label{f:Comp}
\end{figure}

\begin{figure}[!htp]
\begin{center}
\begin{tabular}{ccc}
\includegraphics[width=2.0in]{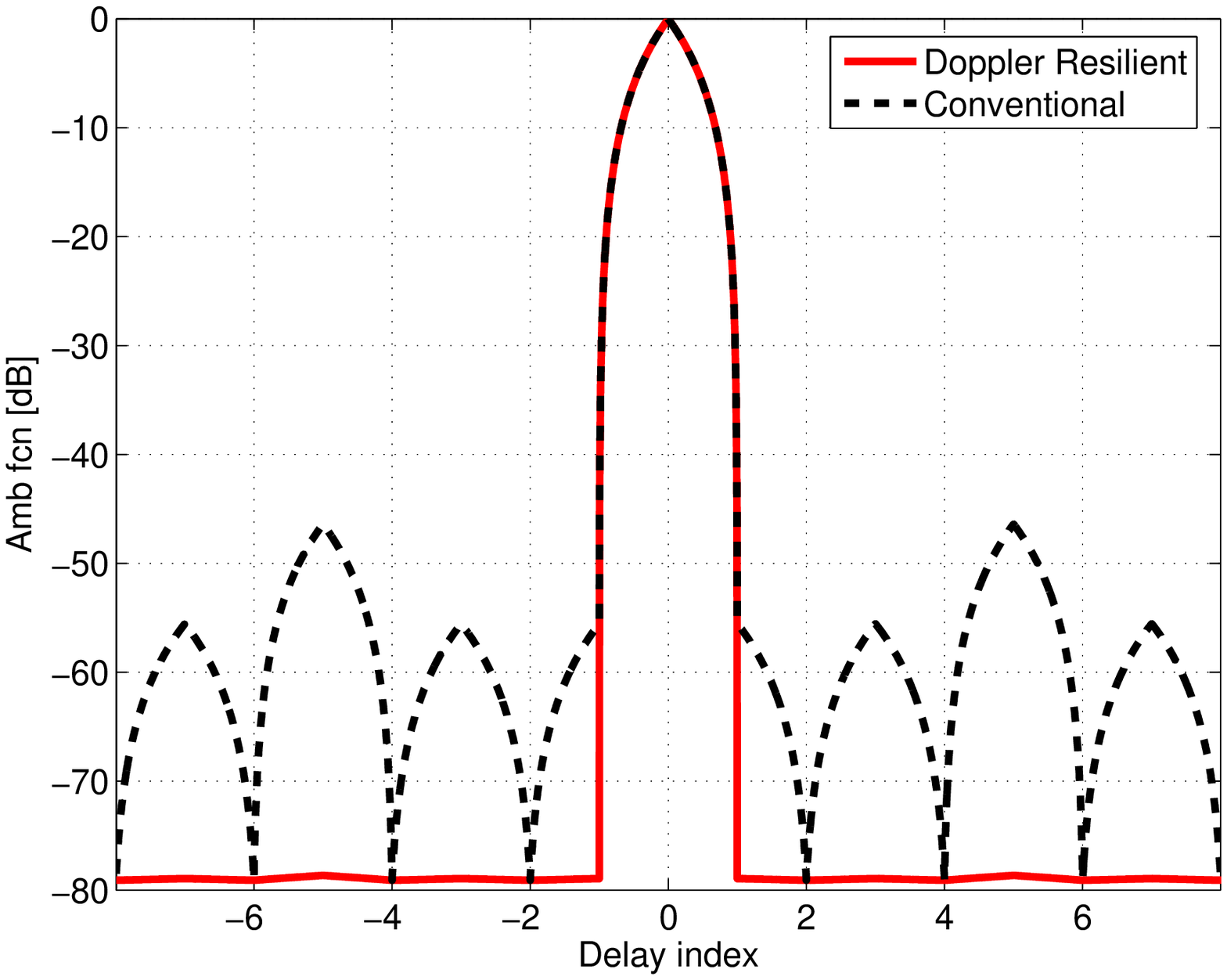}&
\includegraphics[width=2.0in]{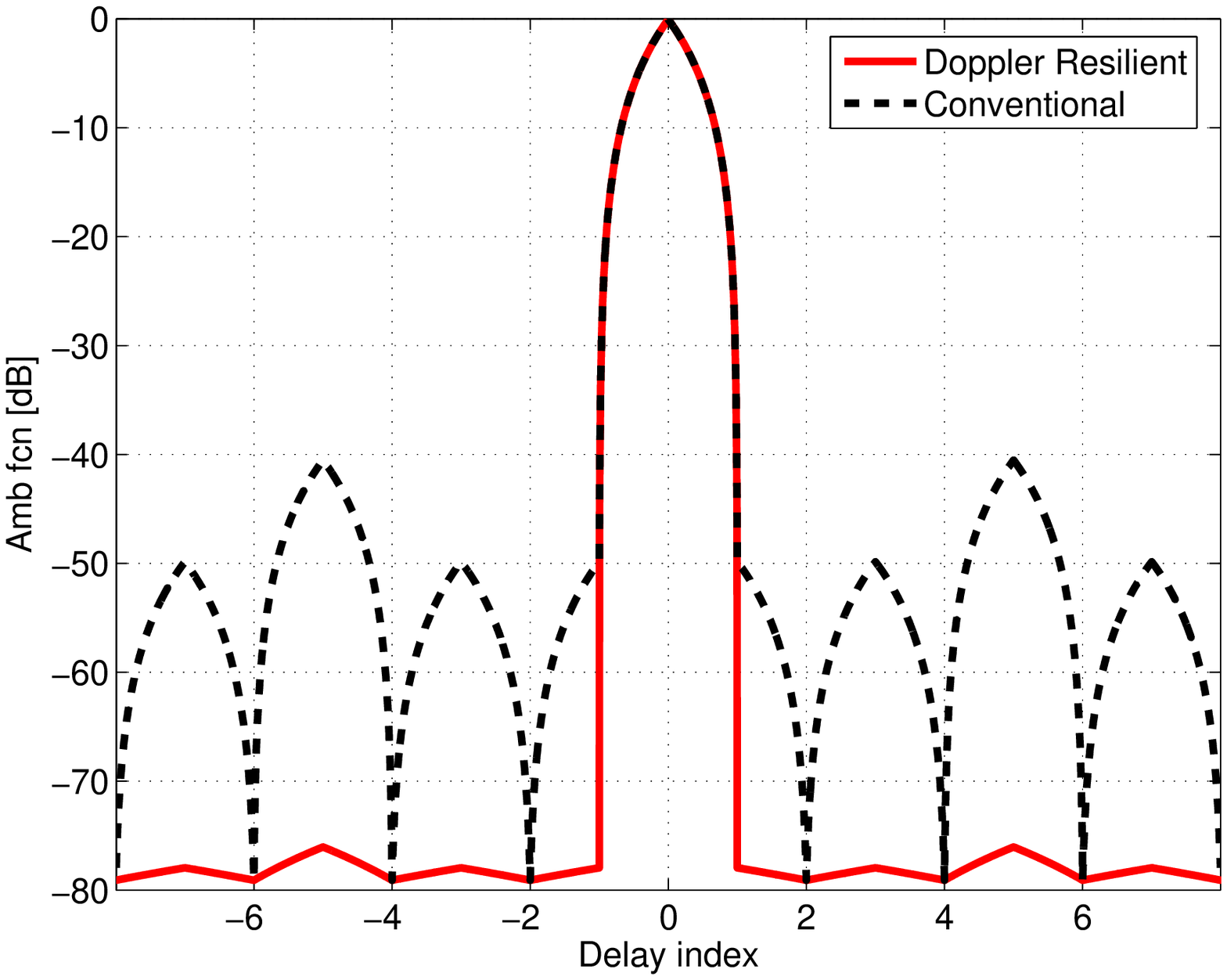}&
\includegraphics[width=2.0in]{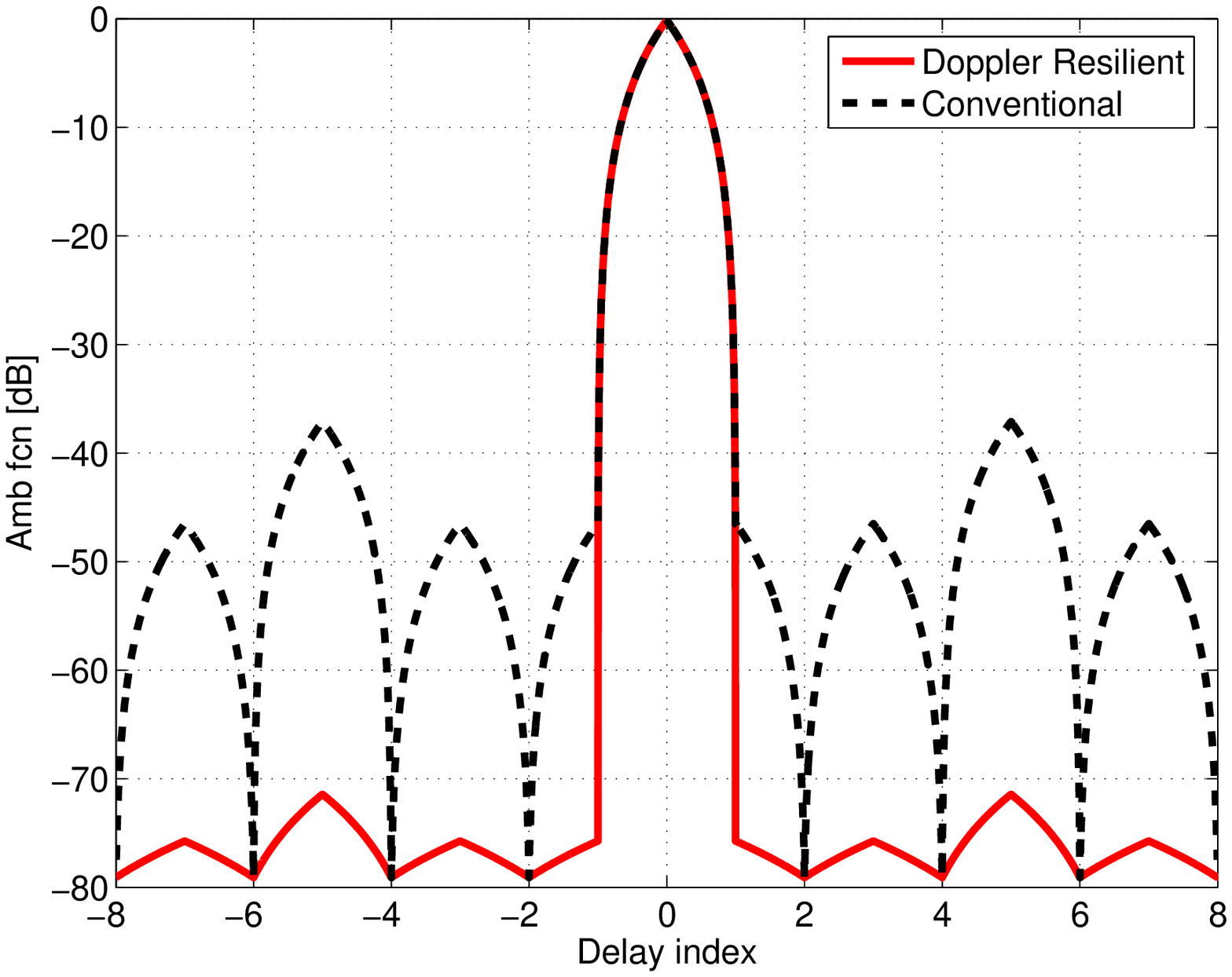}\\
(a) $\theta=0.025$ rad & (b) $\theta=0.05$ rad & (c)
$\theta=0.075$ rad
\end{tabular}
\end{center}
\caption{Comparison of the composite ambiguity functions
$G(z,\theta)$ and $G_c(z,\theta)$ at Doppler shifts (a)
$\theta=0.025$ rad, (b) $\theta=0.05$ rad, and (c) $\theta=0.075$
rad.}\label{f:Comp2}
\end{figure}

Figures \ref{f:Comp}(a),(b) show the plots of the composite
ambiguity functions $G(z,\theta)$ and $G_c(z,\theta)$ versus delay
index $l$ and Doppler shift $\theta$. Comparison of $G(z,\theta)$
and $G_c(z,\theta)$ at Doppler shifts $\theta=0.025$ rad,
$\theta=0.05$ rad, and $\theta=0.075$ rad is provided in Figs.
\ref{f:Comp2}(a)-(c), where the solid lines correspond to
$G(z,\theta)$ (Doppler resilient scheme) and the dashed lines
correspond to $G_c(z,\theta)$ (conventional scheme). We notice
that the peaks of the range sidelobes of $G(z,\theta)$ are at
least $24$ dB (for $\theta=0.025$ rad), $28$ dB (for $\theta=0.05$
rad), and $29$ dB (for $\theta=0.075$ rad) smaller than those of
$G_c(z,\theta)$. These plots clearly show the Doppler resilience
of the waveform vector in \eqref{eq:X1vec}.

\textit{Remark 5:} For a radar with carrier frequency $f_0=2.5$
GHz and PRI$=100$ $\mu$sec, the Doppler shift range of $0$ to
$0.05$ rad ($0.075$ rad) corresponds to a maximum target speed of
$V\approx 35$ kmph ($50$ kmph). To cover a larger speed range we
can use our design with a bank of Doppler filters to provide
Doppler resilience within an interval around the Doppler frequency
associated with each filter.

\subsection{Fully Polarimetric Radar System}

We now consider the matrix-valued composite ambiguity function
$\G(z,\theta)$ in \eqref{eq:CGz}, corresponding to the fully
polarimetric radar system described in Section \ref{sc:2by2}.
Following Theorems 2 and 3, we coordinate the transmission of
eight Golay pairs $(x_0,x_1),\ldots,(x_{14},x_{15})$ across
vertical and horizontal polarizations and over $N=16$ PRIs, so
that in the Taylor expansions of $G_1(z,\theta)$ (the diagonal
element of $\G(z,\theta))$ and $G_2(z,\theta)$ (the off-diagonal
element of $\G(z,\theta))$ the coefficients $C_1(z)$, $C_2(z)$,
and $C_3(z)$ vanish at all nonzero delays and $B_1(z)$, $B_2(z)$,
and $B_3(z)$ vanish at all delays. Letting $\X_0(z)$ and $\X_1(z)$
be the Alamouti matrices in \eqref{eq:X0X1}, then it is easy to
check that the Golay pairs in the following waveform matrix
$\X(z)$ satisfy all the conditions of Theorems 2 and 3:
\begin{equation}\label{eq:X1mat}
\X(z)=\begin{pmatrix} \X_0(z) & \X_1(z) & \X_1(z) & \X_0(z) &
\X_1(z) & \X_0(z) & \X_0(z) & \X_1(z)
\end{pmatrix}.
\end{equation}

\textit{Remark 6:} Representing $\X_0(z)$ and $\X_1(z)$ by $0$ and
$1$ respectively, we notice that the placements of $\X_0(z)$ and
$\X_1(z)$ in $\X(z)$ are determined by the length-$8$ PTM
sequence.

We compare the Doppler resilient transmission scheme in
\eqref{eq:X1mat} with a conventional transmission scheme, where
the Alamouti waveform matrix built from a single Golay pair
$(x_0=x,x_1=y)$ is repeated and the waveform matrix is of the form
\begin{equation}\label{eq:X0mat}
\X_c(z)=\begin{pmatrix} \X_0(z) & \X_0(z) & \X_0(z) & \X_0(z) &
\X_0(z) & \X_0(z) & \X_0(z) & \X_0(z) \end{pmatrix}.
\end{equation}
The matrix-valued composite ambiguity function of $\X_c(z)$ is
given by
\begin{equation}
\G_c(z,\theta)=\X_c(z)\D(\theta)\Xtm_c(z)=\begin{pmatrix}
G_{c1}(z,\theta) & G_{c2}(z,\theta)\\ \widetilde{G}_{c2}(z,\theta)
& G_{c1}(z,\theta)\end{pmatrix},
\end{equation}
where
\begin{equation}
\begin{array}{ll}
G_{c1}(z,\theta)=\|X_0(z)\|^2+e^{j\theta}\|X_1(z)\|^2+\ldots+e^{j(N-2)\theta}\|X_{0}(z)\|^2+e^{j(N-1)\theta}\|X_{1}(z)\|^2
\end{array}
\end{equation}
and
\begin{equation}
\begin{array}{ll}
G_{c2}(z,\theta)=(1-e^{j\theta}+\ldots+e^{j(N-2)\theta}-e^{j(N-1)\theta})X_{0}(z)\Xt_{1}(z).
\end{array}
\end{equation}
The Golay pair $(x,y)$ used in building both $\X(z)$ and $\X_c(z)$
is the length-8 Golay pair in \eqref{eq:basexy}.

We notice that the diagonal elements of $\G(z,\theta)$ and
$\G_c(z,\theta)$, i.e., $G_1(z,\theta)$ and $G_{c1}(z,\theta)$,
are equal to the single channel composite ambiguity functions
$G(z,\theta)$ and $G_c(z,\theta)$, respectively. Therefore, the
plots in Fig. \ref{f:diag} through Fig. \ref{f:Comp2} also apply
for comparing $G_1(z,\theta)$ and $G_{c1}(z,\theta)$. Thus in this
example, we only need to consider the off-diagonal elements of
$\G(z,\theta)$ and $\G_c(z,\theta)$, i.e., $G_{2}(z,\theta)$ and
$G_{c2}(z,\theta)$.

\begin{figure}[htp]
\begin{center}
\begin{tabular}{ccc}
\includegraphics[height=1.3in]{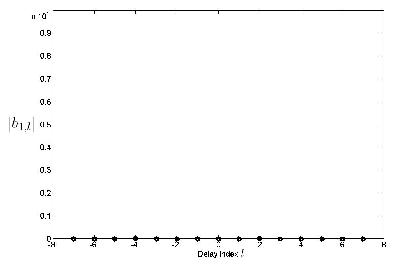}&
\includegraphics[height=1.3in]{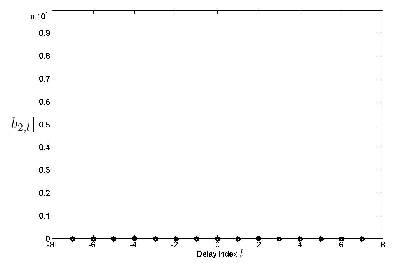}&
\includegraphics[height=1.3in]{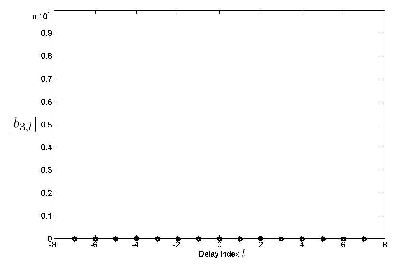}\\
(a) $|b_{1,l}|$ & (b) $|b_{2,l}|$ & (c) $|b_{3,l}|$
\end{tabular}
\end{center}
\caption{The plots of the magnitudes of the coefficients $b_{m,l}$
of two-sided polynomials $B_m(z)=\sum\limits_{l=-(L-1)}^{L-1}
b_{m,l}z^{-l}$, $m=1,2,3$ versus delay index $l$: (a) $|b_{1,l}|$,
(b) $|b_{2,l}|$, and (c) $|b_{3,l}|$.}\label{f:offdiag}
\end{figure}

Referring to the Taylor expansion of $G_2(z,\theta)$ in
\eqref{eq:G2Taylor}, the coefficients $B_1(z)$, $B_2(z)$, and
$B_3(z)$ are each two-sided polynomials of degree $L-1=7$ in
$z^{-1}$ of the form \eqref{eq:Bmpoly}. Figures
\ref{f:offdiag}(a)-(c) show the plots of the magnitudes of the
coefficients $b_{m,l}$, $m=1,2,3$ of these polynomials versus
delay index $l$. We notice that $B_1(z)$, $B_2(z)$, and $B_3(z)$
indeed vanish at all delays.

\begin{figure}[htp]
\begin{center}
\begin{tabular}{cc}
\includegraphics[height=2.0in]{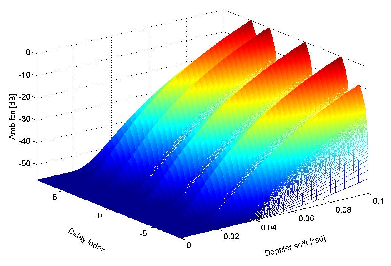}&
\includegraphics[height=2.0in]{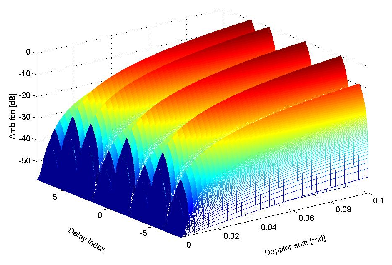}\\
(a) $G_2(z,\theta)$ & (b) $G_{c2}(z,\theta)$
\end{tabular}
\end{center}
\caption{(a) The plot of the off-diagonal element $G_2(z,\theta)$
(corresponding to the Doppler resilient transmission scheme)
versus delay index $l$ and Doppler shift $\theta$, (b) the plot of
the off-diagonal element $G_{c2}(z,\theta)$ (corresponding to the
conventional transmission scheme) versus delay index $l$ and
Doppler shift $\theta$.}\label{f:Comp3}
\end{figure}

\begin{figure}[htp]
\begin{center}
\begin{tabular}{ccc}
\includegraphics[width=2.0in]{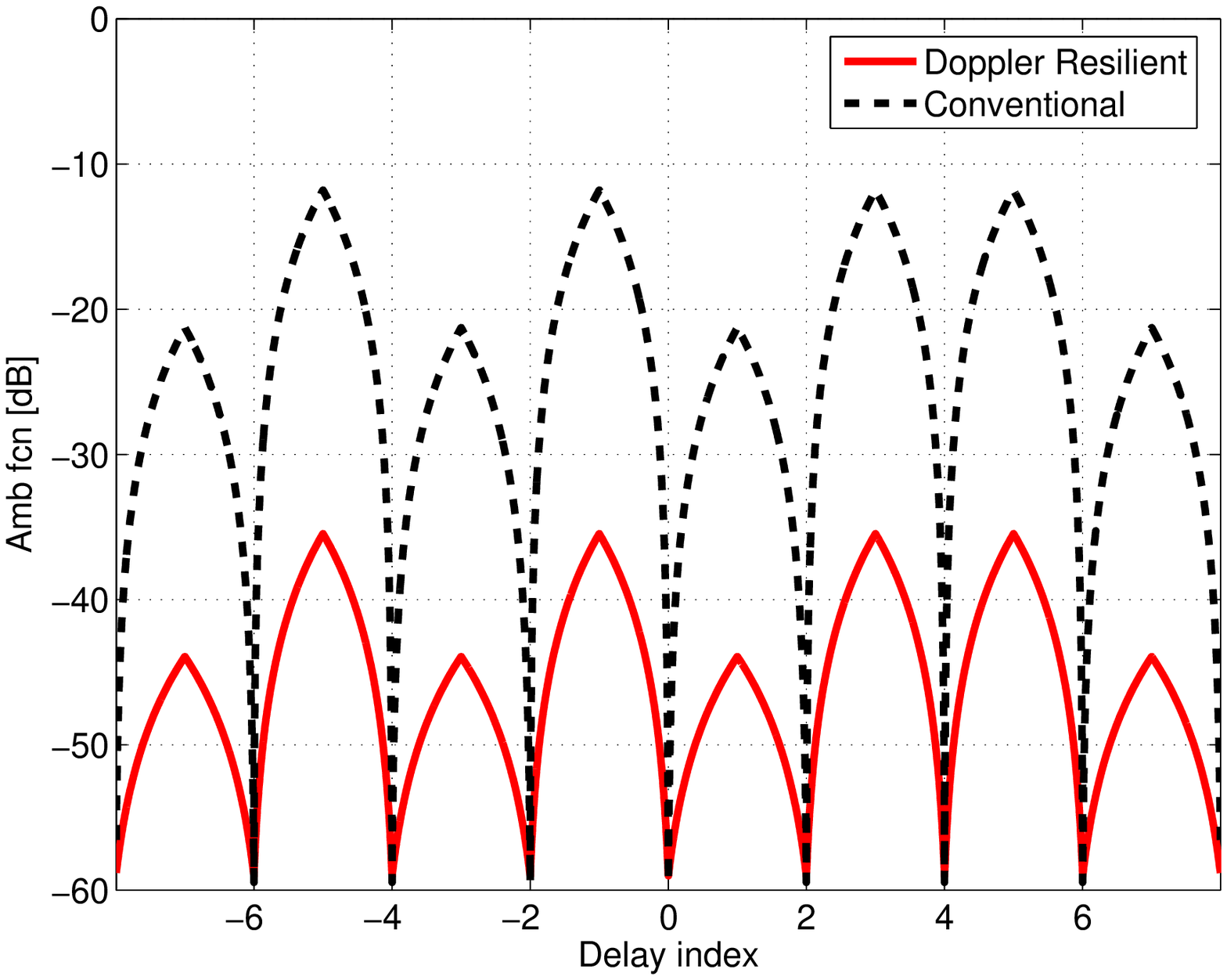}&
\includegraphics[width=2.0in]{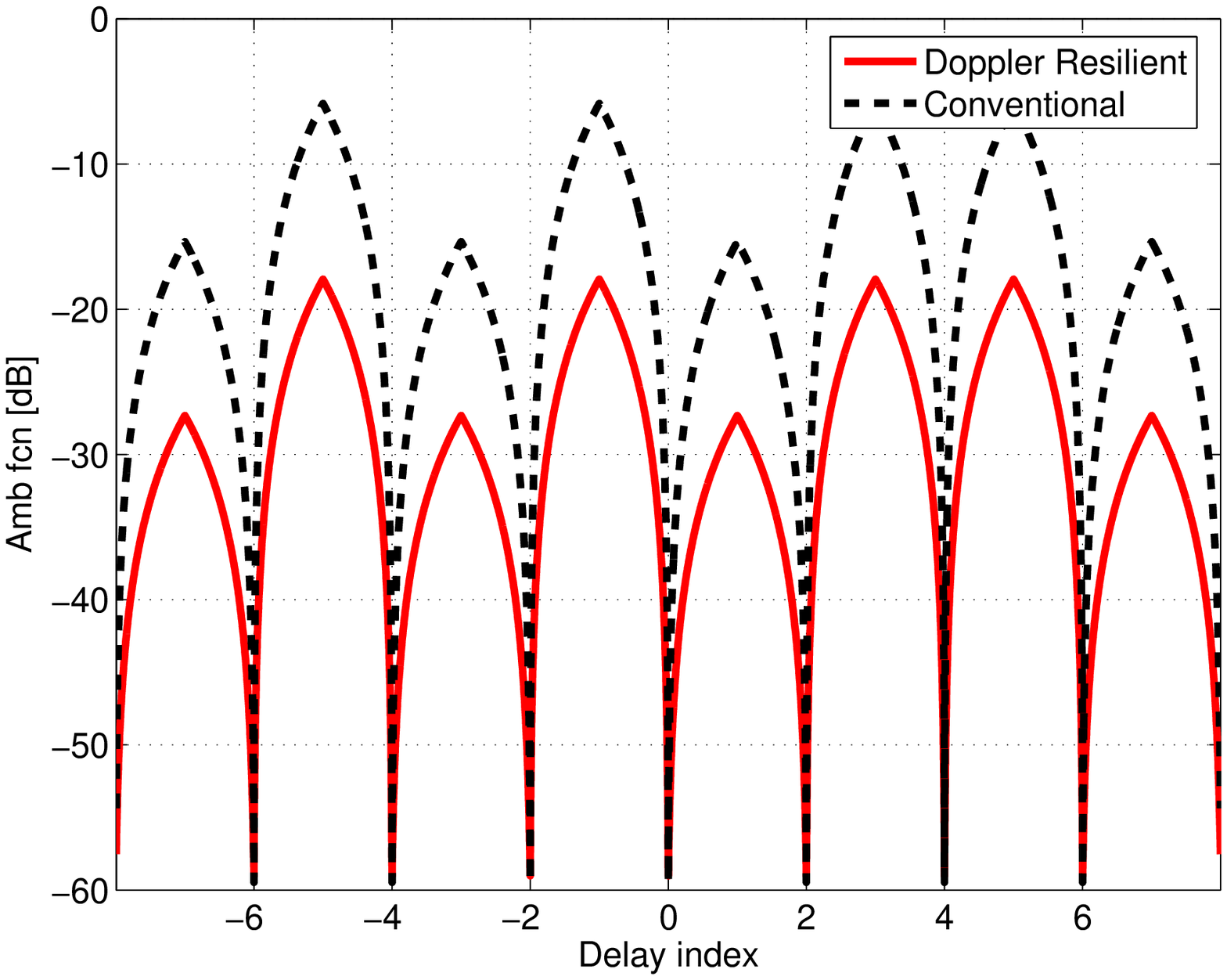}&
\includegraphics[width=2.0in]{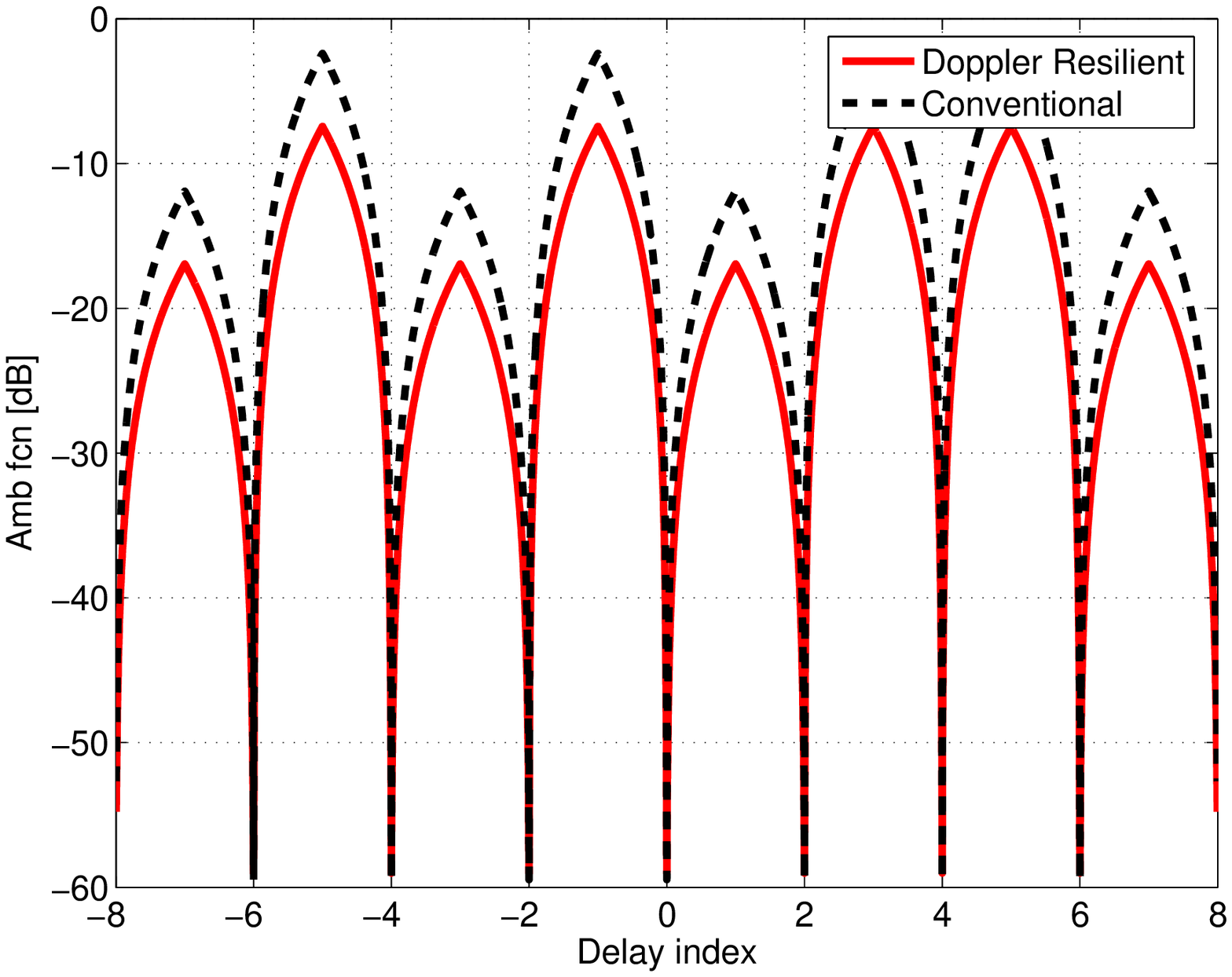}\\
(a) $\theta=0.025$ rad & (b) $\theta=0.05$ rad & (c)
$\theta=0.075$ rad
\end{tabular}
\end{center}
\caption{Comparison of the off-diagonal elements $G_2(z,\theta)$
and $G_{c2}(z,\theta)$ of the matrix-valued ambiguity functions
for the conventional and Doppler resilient schemes at Doppler
shifts (a) $\theta=0.025$ rad, (b) $\theta=0.05$ rad, and (c)
$\theta=0.075$ rad.}\label{f:Comp4}
\end{figure}

Figures \ref{f:Comp3}(a),(b) show the plots of the off-diagonal
elements $G_2(z,\theta)$ and $G_{c2}(z,\theta)$ versus delay index
$l$ and Doppler shift $\theta$. Comparison of $G_2(z,\theta)$ and
$G_{c2}(z,\theta)$ at Doppler shifts $\theta=0.025$ rad,
$\theta=0.05$ rad, and $\theta=0.075$ rad is provided in Figs.
\ref{f:Comp4}(a)-(c), where the solid lines correspond to
$G_2(z,\theta)$ (Doppler resilient scheme) and the dashed lines
correspond to $G_{c2}(z,\theta)$ (conventional scheme). We notice
that the peaks of the range sidelobes of $G_2(z,\theta)$ are at
least $24$ dB (for $\theta=0.025$ rad), $12$ dB (for $\theta=0.05$
rad), and $5$ dB (for $\theta=0.075$ rad) smaller than those of
$G_{c2}(z,\theta)$. These plots together with the plots in Figs.
\ref{f:Comp2}(a)-(c) (corresponding to the diagonal elements of
$\G(z,\theta)$ and $\G_c(z,\theta)$) show the Doppler resilience
of the waveform matrix in \eqref{eq:X1mat}.

\section{Conclusions}

We have constructed a Doppler resilient sequence of Golay
complementary waveforms with perfect autocorrelation at small
Doppler shifts, and extended our results to the design of Doppler
resilient Alamouti waveform matrices of Golay pairs for
instantaneous radar polarimetry. The main contribution is a method
for selecting Golay complementary sequences to force the low-order
terms of the Taylor expansion of a composite ambiguity function
(or Doppler modulated autocorrelation sum) to zero. The
Prouhet-Thue-Morse sequence was found to be the key to selecting
the Doppler resilient Golay pairs. Numerical examples were
presented, demonstrating the perfect correlation properties of
Doppler resilient Golay pairs at small Doppler shifts.

\section*{Acknowledgments}

The authors would like to thank Louis Scharf for his comments on
Section \ref{sc:GPRD}.

\bibliographystyle{IEEEtran}
\bibliography{linalgebbib,DoppRef}

\end{document}